\documentclass[aps,twocolumn,prd,superscriptaddress,letterpaper,nofootinbib,longbibliography]{revtex4-2}
\usepackage{amssymb}
\usepackage{physics}
\usepackage{mathtools}
\usepackage{upgreek}
\usepackage{booktabs}
\AtBeginDocument{
	\heavyrulewidth=.08em
	\lightrulewidth=.05em
	\cmidrulewidth=.03em
	\belowrulesep=.65ex
	\belowbottomsep=0pt
	\aboverulesep=.4ex
	\abovetopsep=0pt
	\cmidrulesep=\doublerulesep
	\cmidrulekern=.5em
	\defaultaddspace=.5em
}
\usepackage{subcaption}

\usepackage{txfonts}
\usepackage{bm}
\usepackage{placeins}
\usepackage{setspace}

\usepackage{tikz}
\usetikzlibrary{calc,positioning,shapes.misc}
\usetikzlibrary{arrows.meta,trees,shadings}
\usetikzlibrary{shapes.geometric}
\usetikzlibrary{intersections}
\newcommand{\laserOneTwo}{LA 12}
\newcommand{\laserOneThree}{LA 13}
\newcommand{\laserTwoOne}{LA 21}
\newcommand{\laserTwoThree}{LA 23}
\newcommand{\laserThreeOne}{LA 31}
\newcommand{\laserThreeTwo}{LA 32}
\newcommand{\spacecraftSeparation}{40mm}
\newcommand{\spacecraftRadius}{11mm}
\newcommand{\fmin}[0]{f_{\mathrm{min}}}
\newcommand{\fmax}[0]{f_{\mathrm{max}}}

\usepackage[colorlinks=true]{hyperref}

\begin{document}

\title{Frequency planning for LISA}

%
%

\author{Gerhard Heinzel}
\affiliation{Max Planck Institute for Gravitational Physics (Albert Einstein Institute), D-30167 Hannover, Germany\\
Leibniz Universit\"at Hannover, D-30167 Hannover, Germany}

\author{Javier \'Alvarez-Vizoso}
\email{javier.vizoso@aei.mpg.de}
\affiliation{Max Planck Institute for Gravitational Physics (Albert Einstein Institute), D-30167 Hannover, Germany\\
Leibniz Universit\"at Hannover, D-30167 Hannover, Germany}

\author{Miguel Dovale-\'Alvarez}
\affiliation{Max Planck Institute for Gravitational Physics (Albert Einstein Institute), D-30167 Hannover, Germany\\
Leibniz Universit\"at Hannover, D-30167 Hannover, Germany}

\author{Karsten Wiesner}
\affiliation{Max Planck Institute for Gravitational Physics (Albert Einstein Institute), D-30167 Hannover, Germany\\
Leibniz Universit\"at Hannover, D-30167 Hannover, Germany}

\begin{abstract}
The Laser Interferometer Space Antenna (LISA) is poised to revolutionize astrophysics and cosmology in the late 2030's by unlocking unprecedented insights into the most energetic and elusive astrophysical phenomena. The mission envisages three spacecraft, each equipped with two lasers, on a triangular constellation with 2.5 million-kilometer arm-lengths. Six inter-spacecraft laser links are established on a laser-transponder configuration, where five of the six lasers are offset-phase-locked to another. The need to determine a suitable set of transponder offset frequencies precisely, given the constraints imposed by the onboard metrology instrument and the orbital dynamics, poses an interesting technical challenge. In this paper we describe an algorithm that solves this problem via quadratic programming. The algorithm can produce concrete frequency plans for a given orbit and transponder configuration, ensuring that all of the critical interferometric signals stay within the desired frequency range throughout the mission lifetime, and enabling LISA to operate in science mode uninterruptedly.
\end{abstract}

\maketitle

\section{Introduction} \label{section:introduction}

The advent of gravitational-wave astronomy has opened a new window onto the cosmos, revealing aspects of the universe that were previously unobservable~\cite{Miller2019}. Ground-based detectors like LIGO~\cite{Collaboration2015} and Virgo~\cite{AdvancedVirgo15} have already made monumental discoveries, detecting the ripples in spacetime caused by merging black holes~\cite{Abbott2016} and neutron stars~\cite{Abbott2017}. However, these terrestrial observatories are limited at low frequency by the seismic noise inherent to Earth and the consequent residual noise of the length and angular control systems needed to keep the detectors in a suitable operating point~\cite{Cahillane2022}.

\begin{figure*}
\centering
\includegraphics[width=\textwidth]{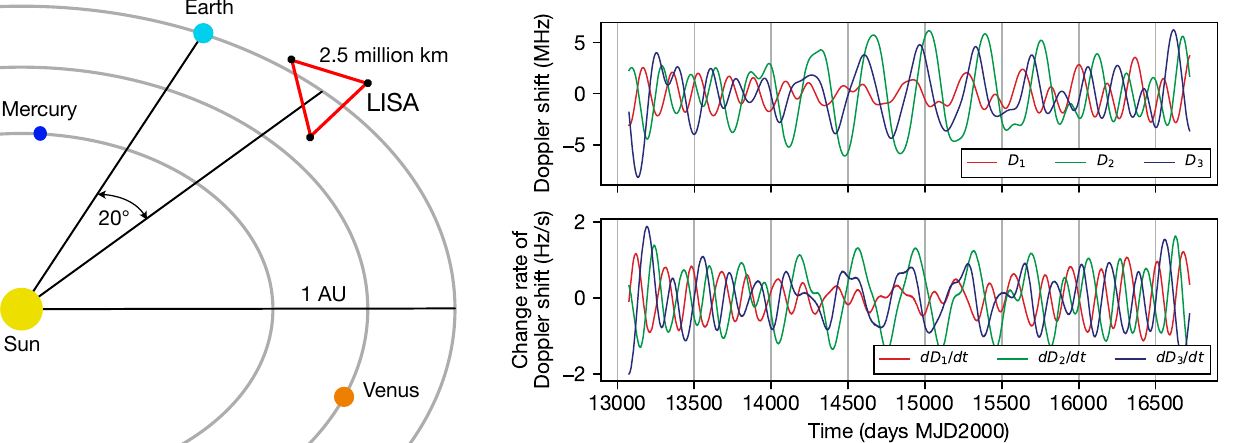}
\caption{Each of the three LISA spacecraft follows an independent heliocentric orbit, trailing behind the Earth by about 20$^{\circ}$, with an inclination of about 1$^{\circ}$ with respect to the ecliptic plane, resulting in a relatively stable triangular formation inclined by 60$^{\circ}$ with respect to the ecliptic. The inter-satellite distances drift at a rate of up to 8\,m/s, yielding frequency shifts of up to $\pm 8\,\rm MHz$ of the laser beams as they travel through the 2.5 million-kilometer arms.}
\label{figure:constellation}
\end{figure*}

The LISA mission aims to put a gravitational-wave detector in space, where these noise sources are marginal, to observe the low frequency part of the gravitational-wave spectrum. Transitioning to space-based detectors like LISA represents a significant leap forward, and promises to unveil a broader spectrum of cosmic events, including supermassive black hole mergers and possibly even echoes from the Big Bang~\cite{Amaro2023}. LISA consists of three identical spacecraft (SC), each forming the corners of a near-equilateral triangle with $2.5\cdot 10^{6}$\,km arm-lengths. The formation orbits the Sun, trailing Earth at a distance of 50 to 65 million kilometers (Figure~\ref{figure:constellation}).

Each spacecraft is equipped with two lasers, each transmitting light to a distant SC and simultaneously producing beatnotes with the incoming received light. The vast distances between the SC and the finite telescope apertures mean that only a tiny fraction of the light transmitted by one SC is captured by the others. Consequently, rather than reflecting the weak incoming light, the system relies on one-way measurements.

Disturbances in the curvature of spacetime caused by gravitational waves change the light travel time, or the optical pathlength, between the spacecraft. These tiny length changes are imprinted into the phases of the aforementioned beatnotes. The beatnote phase shifts (i.e., the one-way measurements) are recorded by the detection system, and can be combined in post-processing along with other data~\cite{Bayle2023} to reveal the gravitational-wave strain using time delay interferometry~\cite{Tinto1999, Tinto2002}.

The orbits of the three spacecraft are carefully designed to maintain the near-equilateral shape of the triangle~\cite{Dhurandhar2005}. However, the relative positions of the spacecraft are not actively controlled and drift at speeds up to 8 meters per second. These drifts result in laser beam frequency variations of up to $\pm 8$\,MHz due to the Doppler effect. A heterodyne interferometric detection system is thus employed, capable of tracking the frequency and phase changes in the MHz beatnotes --- which fluctuate by up to several hertz per second --- with micro-cycle precision over time scales of thousands of seconds.

The detection system electronics are designed to operate in a certain frequency range (e.g., 3 to 30\,MHz). To be able to maintain all of the beatnotes in the desired range, each laser is phase-locked to another one with a programmable offset frequency, with the exception of one laser that is chosen as ``primary'' and locked to an ultra-stable optical reference cavity. Note that, inevitably, some of the locks use light that has travelled along one of the long arms and thus picked up a Doppler shift.

This gives rise to the need to design the so-called \emph{frequency plan}, i.e., the choice of offset frequencies for the five laser transponder locks, for a given orbit and locking configuration. The frequency plan aims to ensure that all beatnote frequencies stay in the desired range for as long as possible, thereby maximizing the uptime of the LISA detector.

Once the laser frequency locking configuration has been decided from one of the feasible constellation designs, linear relationships are established yielding the beatnote frequencies in terms of the Doppler shifts and the offset frequencies at every orbit time. A viable physical solution of this system imposes additional constraints on the beatnotes which translate into linear inequality constraints for the independent offset variables. Posing the problem also requires the selection of 9 sign combinations for the variables, and an additional 12 sign choices for the full set of physical constraints. The final solution is obtained at every time by solving a quadratic optimization problem on the offsets given by minimizing the cost function measuring the distance between the beatnote frequencies and the desired target frequencies. Solving this optimization in a consistent manner for all times of a given orbit yields the desired offset frequencies, and thus, the beatnotes time series that establish the frequency plan.

The paper is structured as follows. Section~\ref{section:method} introduces the constraints on the beatnote frequencies and the frequency plan algorithm. In particular, Subsection~\ref{subsec:geometry_signs} develops a computational-geometric method to obtain an optimal beatnote frequency range so that no interruptions happen in the inter-spacecraft data links during the mission duration; then Subsection~\ref{subsec:optimization} establishes the quadratic optimization algorithm to obtain the frequency plan satisfying all physical constraints. An example laser transponder lock configuration and derived frequency plan is presented in Section~\ref{section:results}. Finally, Section~\ref{section:conclusion} provides a summary and conclusions of this investigation.
 

\section{Method} \label{section:method}

\subsection{Beatnote frequencies and constraints}

The LISA detector is, in essence, a giant unequal-arm Michelson interferometer formed by combining several one-way measurements. Since the inter-SC distances differ by order of $10^8\,$m, the direct combination of the one-way measurements is dominated by an overwhelming coupling of laser frequency noise. A virtual equal-length interferometer, insensitive to laser frequency noise, can be synthesized in post-processing using the technique called TDI~\cite{Tinto1999, Tinto2002}, and a combination of inter-SC absolute ranging~\cite{Esteban2010, Esteban2011} and inter-SC clock synchronization~\cite{Yamamoto2022}.

Each spacecraft contains two optical benches, two independent laser assemblies (LA) and test masses (TM), and a common clock [called ultra-stable oscillator (USO)]. The onboard clock signal is imprinted into the onboard lasers via generation of phase-modulation sidebands. In order for TDI to work, a number of beatnotes are thus generated and recorded via the onboard interferometric detection systems:

\begin{itemize}
\item Six from the six one-way laser links between SC pairs (known as ``inter-satellite interferometers'' or ISI)
\item Six from the interference between the two lasers local to each SC (known as ``reference interferometers'' or RFI)
\item Six from the interference between the two lasers local to each SC, after one of them has reflected off the nearby TM (known as ``test mass interferometers'' or TMI)
\item In addition, a number of sideband-sideband beatnotes are tracked to extract the differential onboard clock signals
\end{itemize}

Since Doppler shifts ($D_1$, $D_2$, and $D_3$, see Figure~\ref{figure:constellation}) are picked up in the inter-SC links, and then coupled from one laser to another via a transponder lock, these highly dynamical signals are present in all of the aforementioned beatnotes. However, a subset of 9 beatnotes (three per SC) is sufficient to fully describe the coupling of the orbital dynamics to the interferometric signals: the six ISI of the constellation, and one RFI per SC. All other beatnotes' orbit-induced dynamics is a copy of one of these 9.

\begin{figure}
\centering
\includegraphics{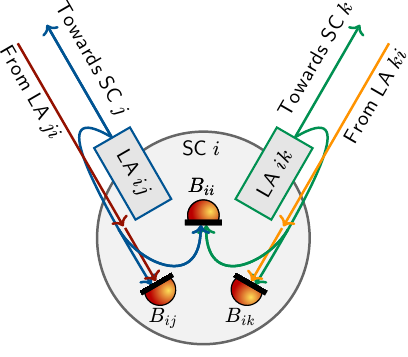}
\caption{Nomenclature used to enumerate spacecraft (SC), laser assemblies (LA), and beatnotes. In SC $i$, LA $ij$ and LA $ik$ are interfered to form the ``local'' beatnote $B_{ii}$. In addition, SC $i$ receives the beams transmitted from the remote LA $ji$ and LA $ki$, which are interfered with the local lasers to form the inter-spacecraft beatnotes $B_{ij}$ and $B_{ik}$ respectively.}
\label{figure:nomenclature}
\end{figure}

Figure~\ref{figure:nomenclature} illustrates the beatnotes in question and introduces the nomenclature used to enumerate spacecraft, lasers, and beatnotes. Spacecraft $i$ contains lasers $ij$ and $ik$, which transmit light to the distant SC $j$ and SC $k$ respectively ($i,j,k \in \{1,2,3\}$). These two lasers are made to interfere locally and produce an RFI beatnote signal at frequency $B_{ii}$. Simultaneously, laser $ij$ is interfered with the light received from its distant counterpart, laser $ji$, producing an ISI beatnote signal at frequency $B_{ij}$. We can thus distinguish between ``local beatnotes'',
\begin{equation}
    B_{11}, B_{22}, B_{33}
\end{equation}
formed by interfering the two onboard lasers, and those that involve a weak beam from a remote spacecraft, which measure the variations of the inter-SC distances, carry the gravitational-wave signal, and have much lower signal-to-noise ratio,
\begin{equation}
    B_{12}, B_{13}, B_{21}, B_{23}, B_{31}, B_{32}.
\end{equation}

Let $L_{ij}$ with $i,j \in \{ 1,2,3\}$ and $i \neq j$ refer to the frequency of laser $ij$ in the constellation of six lasers, and let $D_k(t)$ with $k \in \{ 1,2,3\}$ refer to the Doppler frequency shift acquired by propagating along the long path separating SC $i$ and SC $j$. The beatnote signal frequencies are defined as follows:
\begin{alignat}{2}
B_{11} &= L_{13} - L_{12},        \quad && \text{on SC 1}\nonumber \\
B_{12} &= L_{21} + D_3(t) - L_{12},   \quad && \text{on SC 1}\nonumber \\
B_{13} &= L_{31} + D_2(t) - L_{13},   \quad && \text{on SC 1}\nonumber \\
B_{21} &= L_{12} + D_3(t) - L_{21},   \quad && \text{on SC 2}\nonumber \\
B_{22} &= L_{21} - L_{23},        \quad && \text{on SC 2}\nonumber \\
B_{23} &= L_{32} + D_1(t) - L_{23},   \quad && \text{on SC 2}\nonumber \\
B_{31} &= L_{13} + D_2(t) - L_{31},   \quad && \text{on SC 3}\nonumber \\
B_{32} &= L_{23} + D_1(t) - L_{32},   \quad && \text{on SC 3}\nonumber \\
B_{33} &= L_{32} - L_{31},        \quad && \text{on SC 3},
\label{eq:beatn}
\end{alignat}
One of the six lasers in the constellation is designated as primary and is stabilized in its frequency to a reference cavity. The other five lasers are phase-locked with programmable frequency offsets ($O_1...O_5$) to the primary, either directly or indirectly. Some of these phase locks use light that has been exchanged between satellites, and has thus shifted in frequency due to the Doppler effect. 

The interferometric detection system electronics, tasked with tracking these beatnotes, consists of quadrant photoreceivers~\cite{Barranco2018PhD} and phasemeters~\cite{Heinzel2011, Gerberding2015}. In the quadrant photoreceiver (QPR), an InGaAs quadrant photodiode produces four photocurrent signals that oscillate at the beat frequency ($B_{ij}$) of the two interfering laser beams. Low-noise trans-impedance amplifiers convert the photocurrents into voltage signals and deliver them to the phasemeter. In the phasemeter (of which there is one per inter-SC link), the QPR signals are digitized via analog-to-digital converters (ADC), and processed by field programmable gate arrays (FPGA), implementing digital phase-locked loops (DPLL), which track the frequency and phase of the digitized beatnotes and stream the resulting data to an onboard computer.

The primary reason why frequency planning is necessary is that, due to the technical limitations of the interferometric detection system, the frequencies of all the beatnotes to be detected must fall inside a certain frequency band, [called the heterodyne frequency band, ($\fmin,\fmax$)], whilst undergoing the Doppler-induced frequency drifts (which vary over time and take both positive and negative sign). The following considerations, all pertaining to the interferometric detection system, affect the selection of $\fmin$ and $\fmax$:

\begin{itemize}
    \item The DPLL implemented in the phasemeter FPGA will fail if the beatnote frequency approaches zero (this is the lower bound to $\fmin$), 
    \item The residual intensity noise coupling from the lasers is significant up to a few MHz~\cite{Lennart2023},
    \item The QPR and the analog electronics of the phasemeter have a finite bandwidth, limited by the available components and the capacitance of the photodidoes,
    \item The electronic noise of the QPR increases with frequency due to a combined eﬀect of the input voltage noise of the first stage preamplifier and the photodiode capacitance, 
    \item To prevent aliasing, the beatnote frequencies cannot surpass the Nyquist frequency, which is $1/2$ of the sampling frequency of the ADC (40\,MHz in the current design, this is the upper bound to $\fmax$),
    \item The beatnote frequencies must not cross the frequency of the pilot tone in the phasemeter~\cite{Gerberding2015}, which depending on the internal details of the phasemeter could be at either the lower or at the upper end of the heterodyne frequency band.
\end{itemize}

Moreover, the beatnotes local to each spacecraft must avoid crossing each other in frequency in order to minimize the impact of crosstalk in the phasemeter. The main goal of the frequency plan is deriving a suitable set of the offset frequencies $O_1(t)...O_5(t)$ for each laser lock to ensure that all the beatnotes stay within the ($\fmin,\fmax$) range. The frequency planning therefore depends on:

\begin{itemize}
\item The Doppler shifts experienced in orbit [$D_1(t)$, $D_2(t)$, $D_3(t)$],
\item The choice of primary laser and locking scheme,
\item The choice of $\fmin$ and $\fmax$.
\end{itemize}

For each possible primary laser (out of the six lasers in the constellation), there exist six different locking schemes with varying levels of TDI complexity, resulting in $6^2 = 36$ distinct choices. A computational method to derive and catalog all feasible locking configurations has been developed along with the frequency planning algorithms described in this paper.

The remainder of this section is structured as follows. Subsection~\ref{subsec:math} introduces the mathematical formulation used throughout the rest of the section. Subsection~\ref{subsec:geometry_signs} introduces a computational geometry approach to deriving precise values of $\fmin$ and $\fmax$ guaranteeing the existence of a set of transponder offset frequencies that ensure the uninterrupted operation of the detector throughout the mission lifetime. Then, in Subsection~\ref{subsec:optimization} we present the quadratic optimization algorithm developed to generate concrete frequency plans.

All of the aforementioned computer algorithms (the algorithm used to discover and catalog all possible locking schemes, the one used to derive all feasible values of $\fmin$ and $\fmax$ guaranteeing the existence of a satisfactory frequency plan, and the offset frequency optimizer) have been independently developed in both C and Python. 

\subsection{Mathematical formulation of the constraints on the beatnote frequencies}
\label{subsec:math}

We propose a computational geometry approach to pose the optimization constraints of this problem that will yield an exact solution to the frequency planning, as opposed to other approaches used in the past (such as running optimizers on a set of constant offset frequencies and trying to find sets of frequencies that allow locking for as long as possible before switching to a new set of frequencies~\cite{BarkePhD}).

We combine the Doppler shifts $D_1$, $D_2$, and $D_3$ in a vector $\Vec{D}\in\mathbb{R}^3$, the transponder offset frequencies $O_1$, $O_2$, $O_3$, $O_4$, $O_5$ in a vector $\Vec{O}\in\mathbb{R}^5$, and the beatnote frequencies $B_{11}$, $B_{12}, \dots$, $B_{33}$ in a vector $\Vec{B}\in\mathbb{R}^9$. The laser frequencies $L_{ij}$ are assumed to be positive, but the quantities $D_i$, $O_i$, and $B_{ij}$ can have both signs since they are related to differences between the former.

The choice of locking scheme establishes a linear relationship between the transponder laser frequencies and the primary laser, with the addition of the five offset frequencies. These relations translate at every time into a system of nine linear equations for the beatnote frequencies in terms of the Doppler shifts and offset frequencies, so there are constant matrices $N_1$, $N_2$ (see equation \eqref{eq:m1} below for the values of the example of locking scheme N3-L32), which depend only on the locking configuration and not on the orbit, such that at every time
\begin{align}
\begin{array}{ccccccccc}
\vec{B}(t) & = & N_1 & \cdot & \vec{D}(t) & + & N_2 & \cdot &  \vec{O}(t) \\
9 \times 1 &&
9 \times 3 &&
3 \times 1 &&       
9 \times 5 &&     
5 \times 1 \\ 
\label{eq:quad1}
\end{array}        
\end{align}

Due to the nature of the locking configuration, and thus encoded in the blocks of $N_1$ and $N_2$, the beatnotes can be classified according to whether they are used to lock a transponder laser or not.  In the former case, the corresponding component of $\Vec{B}$ is equal up to a sign to the same component of $\Vec{O}$, and is called a ``locking'' beatnote; the components that are not used in laser locks are called ``non-locking'' beatnotes. The operational distinction between the two sets is that if any of the five locking beatnote frequencies passes through a forbidden frequency range, the respective offset laser lock will temporarily be lost. Depending on where in the locking chain that beatnote is located, the loss of lock might ripple through to other locks. A lost lock means that a re-acquisition procedure must be initiated. A transition through the forbidden region of one of the non-locking beatnotes, on the other hand, will not cause a transponder laser to drop out of lock. In any case, a transition of any beatnote through such forbidden region will cause an interruption in the science data. Such interruption may be short and not significantly affect LISA science. Nevertheless, it is one aim of frequency planning to minimize the number of such interruptions to zero.


Given desired target frequencies for the beatnotes, $\Vec{T}$, with $f_{\min}<|T_i|<f_{\max}$ (e.g., $8$ MHz if that is a sweet spot for noise), if there were no additional constraints, one would optimize for 
\begin{equation}\label{eq:mintarget}
\underset{\Vec{O}\in\mathbb{R}^5}{\operatorname{arg}\operatorname{min}}\; ||\Vec{B}-\Vec{T}||^2
\end{equation}
where $\Vec{B}$ is given by~\eqref{eq:quad1} as a function of $\Vec{O}$ at every time, and thus obtain five offset time series solutions, that in turn yield the beatnote frequency time series. However, the physical constraints on the beatnote frequencies will result in constraints on the offset variables, and problem~\eqref{eq:mintarget} will turn into a quadratic optimization problem with, in fact, linear inequality constraints. There are two types of restrictions on the beatnote frequencies: they must all fall inside the aforementioned frequency band, i.e., 
\begin{equation}\label{eq:bandconstraints}
    f_{\min}\leq |B_{ij}|\leq f_{\max}
\end{equation}
for all $i,j$, and local beatnotes must avoid crossing, i.e., 
\begin{equation}\label{eq:localconstraints}
|B_{ij}|\neq |B_{ii}|,
\end{equation}
for $i\neq j$.

\begin{figure}
\begin{centering}
\includegraphics{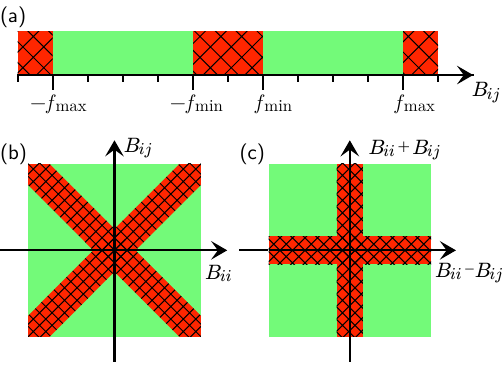}
\caption{ \label{fig:crosslimit} Illustrations of the allowed values of the beatnote frequencies $B_{ij}$ (a) and the constraints to avoid crosstalk (b).}
\end{centering}
\end{figure}

The values $f_{\min}$ and $f_{\max}$ are always positive as they refer to electronically measurable frequencies, i.e., frequencies $-x$ MHz and $+x$ MHz are indistinguishable for the phasemeter. However, in the frequency planning we need to take into account the signs of the differences of laser frequencies and deal with the absolute value of each individual laser frequency (always positive), Doppler shift (positive or negative) and transponder offset frequency (positive or negative). So there are two allowed regions for each frequency difference, and thus beatnote:
\begin{alignat}{2}
-f_{\rm max} &\le B_{ij} &&\le -f_{\rm min}, \quad\text{or} \nonumber\\
f_{\rm min} &\le B_{ij} &&\le f_{\rm max}, \label{eq:fminmax}
\end{alignat}
This condition is illustrated in the diagram of Figure~\ref{fig:crosslimit}a. Hence, we have to consider the possible sign combinations of $\Vec{B}$ and $\Vec{O}$, in particular the signs of the 4 non-locking beatnote components and the signs of the 5 offset components (as the locking components signs are obtained from the latter from at most a fixed sign inversion). We define tuples $\sigma_B$ and $\sigma_O$ of lengths 4 and 5, respectively, with such sign choices, i.e., with elements that can be either `$+1$' or `$-1$':
\begin{align}
\sigma_B &= (\pm 1,\pm 1,\pm 1,\pm 1)\in\Sigma_B, \nonumber\\
\sigma_O &= (\pm 1,\pm 1,\pm 1,\pm 1,\pm 1)\in\Sigma_O,
\end{align}
(note that, despite the notation, each component sign choice is independent of the other components). We are denoting the sets of all such sign combinations as $\Sigma_B,\Sigma_O$, having $2^4=16,\; 2^5 = 32$ elements respectively. A complete sign choice for the frequency band constraints is a tuple of length 9 in a set of $2^9=512$ possibilities.

The additional constraints~\eqref{eq:localconstraints} arise from the desire to avoid the crossing, in both the optical and electronic domains, of the strong RFI or TMI beatnotes, with the weak ISI beatnotes in the same SC that carry much lower signal-to-noise ratio. There is always some crosstalk between phasemeter channels, caused e.g.\ by scattered light or electronic crosstalk in the photoreceivers, harness, or the phasemeter itself. If the amplitude of the offending crosstalking beatnote $B_{ii}$ is smaller than the weak ISI beatnote $B_{ij}$, this would merely be an occasional and predictable glitch in the science data streams. If, on the other hand, the crosstalk of $B_{ii}$ were stronger than $B_{ij}$, it would `hijack' the DPLL tracking loop in the phasemeter as their frequencies cross, and result in important disruptions of the science data stream. This is no fundamental problem, but would set very strict requirements on scattered light and electronic crosstalk to make sure that the crosstalk of $B_{ii}$ is always much weaker than $B_{ij}$ in the phasemeter channels that are supposed to track the latter.

Therefore it is desirable to avoid any such crossing of local beatnotes. In other words, we would like that equations~\eqref{eq:localconstraints} are satisfied at each spacecraft, yielding in total six inequality conditions in the constellation. Anticipating the geometric considerations of the following section, we convert these into linear inequalities, illustrated here for a fixed case of $i\neq j$, $|B_{ij}|\neq|B_{ii}|$ (one of six, three possibilities for $i$ and two for $j$). In order to have some margin, e.g.\ for the GHz clock sidebands and PRN modulation, we demand for some $\varepsilon>0$
\begin{align}
\left|\,|B_{ij}|-|B_{ii}|\,\right|>\varepsilon.
\label{eq:bndiff}
\end{align}
This condition is illustrated in the diagram of Figure~\ref{fig:crosslimit}b. It can more conveniently be expressed in terms of $B_{ii}-B_{ij}$ and $B_{ii}+B_{ij}$, as illustrated in the diagram of Figure~\ref{fig:crosslimit}c. 
We therefore have four separate allowed regions per case $i\neq j$:
\begin{equation}\label{eq:constraints_crossing_explicit}
\begin{array}{ccrcccr}
B_{ij}-B_{ii} & > & \varepsilon & \text{and} & B_{ij}+B_{ii}& > & \varepsilon, \quad\text{or}\\
B_{ij}-B_{ii} & > & \varepsilon & \text{and} & B_{ij}+B_{ii}& < & -\varepsilon, \quad\text{or}\\
B_{ij}-B_{ii} & < & -\varepsilon & \text{and} & B_{ij}+B_{ii}& > & \varepsilon, \quad\text{or}\\
B_{ij}-B_{ii} & < & -\varepsilon & \text{and} & B_{ij}+B_{ii}& < & -\varepsilon. \phantom{\quad\text{or}}
\end{array}
\end{equation}
Since there are six such beatnote differences in the constellation, we arrive at $4^6=4096$ sign combinations, which we denote by
\begin{align}
\sigma_C=(\underbrace{\pm1, \dots, \pm1}_{12\times})\in\Sigma_C.
\label{eq:sigmac}
\end{align}
Notice that for a given fixed sign choice of the tuples $\sigma_B, \sigma_O$, only some of the above inequalities are consistent, and in turn only certain combinations of $\sigma_C$ are feasible: for example, if $B_{ij}>0$ and $B_{ii}<0$, then only the choice of $B_{ij}-B_{ii}>\varepsilon$ is consistent, reducing the number of feasible sign combinations for this specific crossing constraint from 4 to 2.


\subsection{Computational geometry algorithm to decide an optimal heterodyne frequency band ($\fmin, \fmax$)}
\label{subsec:geometry_signs}

The first step towards solving the minimization~\eqref{eq:mintarget} subject to inequalities~\eqref{eq:bandconstraints} and~\eqref{eq:bndiff} is determining which sign choices $\sigma_B, \sigma_O$ are feasible, in the sense of allowing a set of permissible offset frequencies such that the optimization problem has a solution, and of those feasible ones which one is more appropriate in some technical sense. This must be done at a fixed time $t$, such that the Doppler shifts are given and fixed. The separate results for each $t$ can then be combined into solutions for the whole time interval (e.g., the whole mission duration).

Since the value and sign of the locking beatnotes is fixed up to sign inversion by the offsets, Equation \eqref{eq:mintarget} can be considered only for the vector component formed by the 4 non-locking beatnotes, which by abuse of notation we denote from now on also as $\Vec{B}\in\mathbb{R}^4$:
\begin{align}
\begin{array}{ccccccccc}
\vec{B}(t) & = & M_1 & \cdot & \vec{D}(t) & + & M_2 & \cdot &  \vec{O}(t). \\
4 \times 1 &&
4 \times 3 &&
3 \times 1 &&       
4 \times 5 &&     
5 \times 1 \\ 
\label{eq:exact1}
\end{array}        
\end{align}
Here $M_1, M_2$ are the corresponding matrix blocks from Equation~\eqref{eq:quad1} when reordering the components into locking and non-locking variables (see the Results section for an explicit example). These matrices are constant and given by the locking configuration, whereas the Doppler shifts are time series computed from the orbits. The task at hand is then to find sign choices $\sigma_B,\sigma_O$ such that there are offsets $\Vec{O}$ and non-locking beatnotes $\Vec{B}$ both satisfying the constraints~\eqref{eq:bandconstraints} and Equation~\eqref{eq:exact1}. This can be solved by computational-geometrical methods for a fixed time $t$. The aim is to find consistent and fixed sign choices for as long a period of time as possible to minimize the need to switch signs in order to avoid brief interruptions of the science data streams and some inter-spacecraft data links.
\begin{figure*}[t!]
\begin{subfigure}{0.49\linewidth}
\centering
\includegraphics{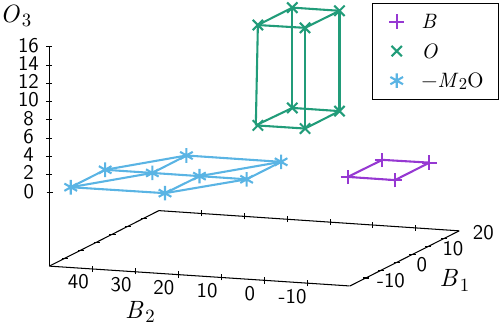}
\caption{Minkowski sum example in reduced dimension.}
\label{fig:minkowski1}
\end{subfigure}\hfill
\begin{subfigure}{0.49\linewidth}
\centering
\includegraphics{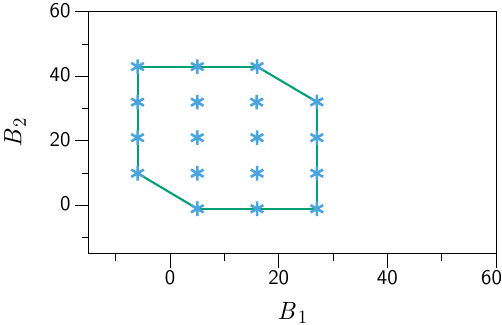}
\caption{Convex hull example in reduced dimension. }
\label{fig:minkowski2}
\end{subfigure}\hfill
\caption{Illustration of the geometrical procedure in reduced dimensions. In (a) edges and corners are shown for the sets $\widehat{B}$ (2-D), $\widehat{O}$ (3-D) and $-M_2\widehat{O}$ (projected in 2-D). In (b) the points are all possible sums of one corner from $\widehat{B}$ plus one corner from $-M_2\widehat{O}$. The line represents the convex hull as computed by {\tt qhull}, which is the Minkowski sum $\widehat{B} \oplus (-M_2\widehat{O})$. The test whether a solution exists can now be performed by checking if $M_1\vec{D}$ falls in that convex hull.}
\label{fig:Minkowski_example}
\end{figure*}
Rearranging the previous equation, we are given a fixed vector $M_1\cdot\Vec{D}(t)\in\mathbb{R}^4$ that must decompose as
\begin{align}
\begin{array}{ccccccccc}
M_1 & \cdot & \vec{D}(t) & = & \vec{B} & - & M_2 & \cdot &  \vec{O} \\
4 \times 3 &&
3 \times 1 &&       
4 \times 1 &&
4 \times 5 &&     
5 \times 1 \\ 
\label{eq:exact2}
\end{array}        
\end{align}
for some non-locking beatnotes and offsets that must lie in their respective domains of feasibility given by the constraints~\eqref{eq:bandconstraints}, which we denote by $\widehat{B}\subset\mathbb{R}^4$ and $\widehat{O}\subset\mathbb{R}^5$ respectively. For a fixed $\sigma_B$, the domain of feasible beatnote frequencies $\widehat{B}$ is a 4D hypercube, as it is the Cartesian product of closed intervals $[f_{\min}, f_{\max}]$ or $[-f_{\max}, -f_{\min}]$ for each bounded beatnote component. Likewise, the corresponding offset domain $\widehat{O}$ is a 5D hypercube. These sets are convex and depend on the specific $\sigma_B,\sigma_O$ chosen. The image of a hypercube by a linear transformation is a polytope (i.e., a finite hypervolume limited by hyperplanes), and the linear image of a convex polytope is known to be still convex, so $-M_2\cdot\widehat{O}\subset\mathbb{R}^4$ is a convex polytope. Therefore for a feasible solution to exist we require that there exist vectors $\Vec{x}$ and $\Vec{y}$ such that
\begin{align}
M_1  \cdot  \vec{D}(t)  = \vec{x} + \vec{y} \quad\text{with}\quad  \vec{x} \in \widehat{B}, \quad \vec{y}\in -M_2\cdot\widehat{O}
\label{eq:min1}
\end{align}
in 4D space.
The next key step is to write this as:
\begin{align}
M_1  \cdot  \vec{D}(t)  \quad\in\quad  \widehat{B} \oplus (-M_2\cdot\widehat{O}),
\label{eq:min2}
\end{align}
where $\oplus$ represents the Minkowski sum of two sets, which is by definition the set given by all the possible linear combinations of the vectors from each set~\cite{Schneider2013}. This procedure is useful because the Minkowski sum of two convex polytopes is again a convex polytope, and this can be determined efficiently since a fast test exists to verify whether or not any given 4D point falls inside. As the Minkowski sum depends only on the sign choices $\sigma_B, \sigma_O$, the locking scheme, and the heterodyne frequency band ($\fmin,\fmax$), the complete time series $M_1\cdot\Vec{D}(t)$ can be tested at all times to check whether the same sign choices are feasible and constant throughout the orbit. In particular, a minimal unique representation of a convex polytope is given by the convex hull of a list of points on its boundary or in its interior, if that list contains all corner points.

The detailed procedure to construct $\widehat{B}\oplus(-M_2\widehat{O})$ is as follows: 
\begin{enumerate}
    \item Given $\sigma_B$, the vertices of the hypercube $\widehat{B}$ are completely determined by $(\pm f_1, \pm f_2, \pm f_3, \pm f_4)$ where $f_i\in\{f_{\min},f_{\max}\}, i=1,\dots, 4$, are chosen in all possible combinations, and the signs are fixed by $\sigma_B$, yielding $16$ corner points. Similarly for $\widehat{O}$ given $\sigma_O$, yielding $32$ corner points.

    \item Compute $\vec{y} = -M_2\cdot\vec{O}$ for every corner point $\vec{O}$ of $\widehat{O}$.
    \item Collect the points $\vec{B}+\vec{y}$, for every corner point $\vec{B}$ of $\widehat{B}$, as candidates for a corner of the Minkowksi sum, obtaining $512$ points of $\widehat{B}\oplus(-M_2\widehat{O}$).
    \item Construct the convex hull of the $512$ points to eliminate interior points and boundary points that are not corner points.
\end{enumerate}
A visual example in lower dimension is shown in Figure \ref{fig:Minkowski_example}, where the beatnote range hypercube is represented by a two-dimensional square on the plane (in purple) and the offset hypercube by a three dimensional orthohedron in space (in green). The linear map given by matrix $-M_2$ projects down the latter into the two-dimensional plane, where some of the corner points may become internal (in blue), thus requiring the computation of the convex hull to extract a representation of the vertices of the convex polytope $-M_2\cdot\widehat O$.

Software packages like \verb|qhull|~\cite{Barber1996} provide tools to perform these operations, and moreover provide the normal equations (in the form of coefficients $a_1, a_2, a_3, a_4, b$) of the hyperplane faces that define the outer sides of the convex hull, such as
\begin{equation}\label{eq:polytope_face}
    -a_1X_1 - a_2X_2 - a_3X_3 - a_4X_4 - b = 0,
\end{equation}
so that a point $\vec{X}\in\mathbb{R}^4$ is on the boundary when the left-hand side is zero, and inside the polytope if it is positive.

With this, the test to check whether $M_1\cdot\vec{D}(t)\in\mathbb{R}^4$ lies inside the feasibility region $\widehat{B} \oplus (-M_2\cdot\widehat{O})$ consists of simply substituting $\vec{X}(t)=M_1\cdot\vec{D}(t)$ in Equation \eqref{eq:polytope_face} for all faces of the Minkowski sum. If any result is negative, the point lies outside the convex hull, and thus for that particular $\vec{D}(t)$ no solution exists for the given sign choices $\sigma_B, \sigma_O$ and frequency range $f_{\min},f_{\max}$. We can encapsulate this information in the margin function
\begin{equation}\label{eq:margin}
    m(t,\sigma_B,\sigma_O) = \underset{i}{\operatorname{min}}\; (-a_{1,i}X_1 - a_{2,i}X_2 - a_{3,i}X_3 - a_{4,i}X_4 - b_i = 0),
\end{equation}
(the dependence on $f_{\min}, f_{\max}$ is left implicit), where $i$ runs over the hyperplane equations that define the faces of the convex hull, and $\vec{X}(t)=M_1\cdot\vec{D}(t)$. This minimum value represents, in MHz, how far away one is from hitting a boundary of the feasibility region, where its sign represents whether the point is inside (positive) or outside (negative) of the constraining polytope. This margin function can be employed to check all $512$ sign combinations $\sigma_B,\sigma_O$ at multiple times $t$, typically for the whole mission duration. Define margin functions $m_1,m_2,m_3$ which only depend on the orbit, the locking scheme, and $\fmin,\fmax$, as:

\begin{align}
    & m_1 =  \underset{\sigma_O}{\operatorname{max}}\,\underset{\sigma_B}{\operatorname{max}}\,\underset{t}{\operatorname{min}}\; m(t,\sigma_B,\sigma_O),\label{eq:margins} \\
    & m_2 =  \underset{\sigma_O}{\operatorname{max}}\,\underset{t}{\operatorname{min}}\,\underset{\sigma_B}{\operatorname{max}}\; m(t,\sigma_B,\sigma_O), \\
    & m_3 =  \underset{t}{\operatorname{min}}\,\underset{\sigma_O}{\operatorname{max}}\,\underset{\sigma_B}{\operatorname{max}}\; m(t,\sigma_B,\sigma_O),
\end{align}
which fulfill $m_1\leq m_2 \leq m_3$. The meaning of these functions is as follows: if $m_1$ is positive there is at least one sign combination which permits locking at all times without sign switching, i.e., the optimal situation. If $m_3$ is positive, the constellation can be locked at all times, but possibly sign switching is necessary at some point of the mission. If $m_3$ is negative, there is no feasible solution for at least one time $t$ (i.e., the specific combination of orbit, locking scheme, and $f_{\min}, f_{\max}$ is unacceptable since there are times at which some locks are not feasible). If $m_2$ is positive the sign switches happen for the non-locking beatnotes only, i.e., $\sigma_O$ can be held constant for the whole mission duration but $\sigma_B$ cannot, which means that although the laser locks can be maintained without interruption, there would still be brief interruptions of some inter-spacecraft data links and therefore disruptions of the science data stream.

For a given orbit, locking scheme, and heterodyne frequency range, there are four cases to distinguish:
\begin{enumerate}
    \item[\bf Case 0:] $m_1\leq m_2 \leq m_3 < 0$: At some time $t$ no solution exists at all.
    \item[\bf Case 1:] $m_1\leq m_2 \leq 0 < m_3$: A solution exists for all times requiring sign switching of both locking and non-locking beatnotes.
    \item[\bf Case 2:] $m_1\leq 0 < m_2 \leq m_3$: A solution exists for all times requiring sign switching of only the non-locking beatnotes.
    \item[\bf Case 3:] $0 < m_1\leq m_2 \leq m_3$: A solution exists for all times requiring no sign switching at all.
\end{enumerate}
The conditions $m_1=0$, $m_2=0$ and $m_3=0$ define boundary contours between these cases in the $(\fmin,\fmax)$-plane for a given orbit and locking scheme. These functions thus provide a method to choose the values of $\fmin$ and $\fmax$ such that the corresponding point lies in the region of Case 3, and no sign switching is necessary at any point of the mission. Note however that these refer to the carrier-carrier beatnotes, an extra margin of 1 to 2\,MHz should be foreseen for the clock sidebands. Moreover, the pilot tone must be taken into account, and adding some more margin to account for potential variations of the orbits is advisable.

The previous method does not take into account the additional constraints of Equations \eqref{eq:localconstraints}, since then the set of feasible offsets $\widehat{O}$ is no longer convex. Hence, it does not determine the signs $\sigma_C$. This is however not a limitation since, in practice, for parameter sets that allow a solution without the extra constraints, solutions also exist satisfying the extra constraints (at most a small extension of the frequency range is needed). This is due to the fact that imposing the additional linear inequality constraints for fixed signs $\sigma_C$ restricts the feasibility domain to a subdomain inside the already discussed convex polytope. Therefore, the election of the signs $\sigma_B, \sigma_O$ and the fixing of $f_{\min}, f_{\max}$ in Case 3 by the computational-geometric method above are necessary, even if not sufficient, to obtain optimal solutions. Informed by these choices the quadratic programming algorithm explained in the next subsection finds initial conditions and feasible optimal solutions satisfying all constraints with a sign combination fixed throughout the entire time series, and reports when no feasible solution exists. Furthermore, it takes advantage of exploring the possible consistent signs $\sigma_C$ (a subset out of the 4096 cases) in order to get optimal solution time series that maximize the margin function or any other metric, e.g., minimizing the beatnote rate of change RMS (of interest to TDI operations).

\begin{figure}[t!]
    \centering
    \includegraphics{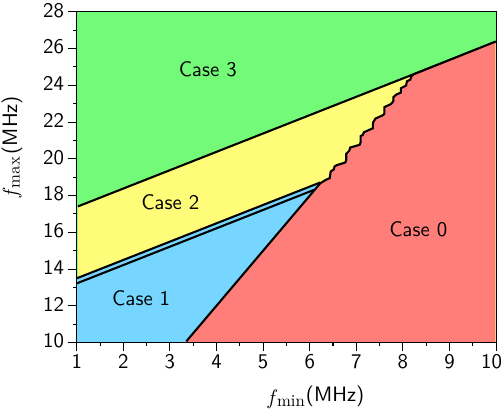}
    \caption{Contour plots of the feasible frequency ranges for non-swap configurations, using ESA orbits \texttt{`op25\_340\_10y3t.xyz'} (10 years). The boundary between case 0 and case 1 as well as between case 2 and case 3 (bold lines) is independent of the configuration, whereas the boundary between case 1 and case 2 (thin lines) depends on the configuration and choice of primary laser.}
    \label{fig:contours}
\end{figure}

\subsection{Offset frequency planning optimization}
\label{subsec:optimization}

By the previous discussion we can choose a suitable frequency range $(f_{\min}, f_{\max})$ and sign combination $(\sigma_B,\sigma_O)$, ideally constant for the whole mission duration if the frequency range is in the region of Case 3, that guarantees the existence of feasible offset frequency solutions. In order to find the actual values of these offset frequencies $\vec{O}(t)$ we propose a geometric optimization where we minimize the Euclidean distance between the corresponding beatnote vectors $\vec{B}(t)$ and target frequencies $\vec{T}$ of our choice, typically optimal values in the detection chain sensitivity. The figure of merit is then minimizing the norm of $\vec{B}-\vec{T}$ by varying $\vec{O}$ through the dependency of Equation \eqref{eq:quad1}, which rearranging terms and squaring yields
\begin{equation}\label{eq:optimization_posed}
    || N_2\cdot\vec{O}(t) - (\vec{T} - N_1\cdot\vec{D}(t)) ||^2 \longrightarrow \operatorname{min},
\end{equation}
i.e., we are minimizing a polynomial of second order in five variables, the $\vec{O}$ components, a typical problem in quadratic programming of the form 
$$
\underset{\vec{x}}{\operatorname{arg}\,\operatorname{min}}\; ||A\cdot\vec{x}-\vec{b}||^2,
$$
with a rectangular matrix $A=N_2$, vector $\vec{b}=\vec{T} - N_1\cdot\vec{D}(t)$, and linear inequality constraints on $\vec{x}=\vec{O}(t)\in\mathbb{R}^5$.

In order to specify the constraints we need to rewrite Equations \eqref{eq:fminmax}, which state the beatnotes must fall within the allocated frequency band, and \eqref{eq:constraints_crossing_explicit}, which prevent local crossing of beatnotes within a spacecraft, in terms of the offsets frequencies using \eqref{eq:quad1}. Optimal solutions can be found by applying quadratic programming techniques when the sign is fixed per beatnote, which justifies the method of the previous section for finding a suitable $\sigma = (\sigma_O,\sigma_B)$ in advance. Recall that $\sigma_B$ determines the sign of the four non-locking beatnotes, whereas the other five locking beatnote frequencies are equal to the offset frequencies up to a sign, encoded in the respective entries of the matrix $N_2$ which is given by the constellation configuration, so that their sign is completely determined by this matrix and $\sigma_O$. Thus, for a particular $\sigma$ and locking scheme, there are range limits for the offsets (and in turn for the locking beatnotes) and non-locking beatnotes,
\begin{align}
\min{}_i \le O_i \le \max{}_i, \quad i=1 \dots 5,\nonumber \\
\min{}_j \le B_j \le \max{}_j, \quad j=1 \dots 4,
\label{eq:bminmax}
\end{align}
where $B_j$ are the non-locking beatnotes as in Equation \eqref{eq:exact1}, and either
\begin{align}
\min{}_i &= -f_{\rm max},\nonumber\\
\max{}_i &= -f_{\rm min},
\end{align}
if the corresponding entry in $\sigma_O$ is negative, or
\begin{align}
\min{}_i &= f_{\rm min},\nonumber\\
\max{}_i &= f_{\rm max},
\end{align}
otherwise. Analogously for $\min{}_j$ and $\max{}_j$ from $\sigma_B$. The nine inequalities~(\ref{eq:bminmax}) translate into 18 inequalities that a quadratic programming
algorithm can handle:
\begin{alignat}{2}
O_i &\ge &&\min{}_i,\nonumber\\
(O_i \le \max{}_i) \quad\rightarrow\quad -O_i &\ge -&&\max{}_i.\nonumber \\
B_j &\ge &&\min{}_j,\nonumber\\
(B_j \le \max{}_j) \quad\rightarrow\quad -B_j &\ge -&&\max{}_j.
\end{alignat}
Because of the dependencies of Equation~(\ref{eq:exact1}), we obtain the final 18 inequality constraints on the offset frequencies as independent variables of the optimization method:
\begin{alignat}{2}
O_i &\ge &&\min{}_i,\nonumber\\
-O_i &\ge -&&\max{}_i.\nonumber \\
(M_2 \cdot \vec{O})_j &\ge &&\min{}_j - (M_1 \cdot  \vec{D}(t))_j,\nonumber\\
-(M_2 \cdot \vec{O})_j &\ge - &&\max{}_j + (M_1 \cdot  \vec{D}(t))_j .\label{eq:quadprog:con18}
\end{alignat}
all of which must be fulfilled simultaneously to minimize \eqref{eq:optimization_posed}.

In addition, the 12 constraints \eqref{eq:constraints_crossing_explicit} to avoid beatnote local crossing must be included. There are 4096 sign combinations, encoded in $\sigma_C$, and the smaller subset of consistent combinations with respect to the fixed $(\sigma_O, \sigma_B)$ must be exhaustively explored in order to select the best optimization result. Let the 12 possible $B_{ij}-B_{ii}$ and $B_{ij}+B_{ii}$, denoted below as $\Delta B_{k}, k=1,\dots, 12$, be written in terms of the offsets as
\begin{equation}\label{eq:DB_SB}
    \Delta B_k = \sum_{i=1}^3 (S_1)_{ki}D_i + \sum_{i=1}^5 (S_2)_{ki}O_i,
\end{equation}
where the matrices $S_1, S_2$ are computed from $N_1, N_2$ via substitution of Eq. (\ref{eq:exact1}) in $B_{ij}-B_{ii}$ and $B_{ij}+B_{ii}$. Then the additional constraints are
\begin{align}
& \Delta B_k \le-\varepsilon,\quad\text{if}\quad(\sigma_C)_k<0,\quad\text{or}\nonumber\\
& \Delta B_k \ge\varepsilon,\quad\text{if}\quad(\sigma_C)_k>0,\quad
k=1 \dots 12.\nonumber    
\end{align}
Rearranging terms, we get the final 12 linear constraint inequalities on the offset frequencies as independent variables that must be added to \eqref{eq:quadprog:con18}:
\begin{align}
-\sum_{i=1}^5 (S_2)_{ki}O_i &\ge\varepsilon+\sum_{i=1}^3 (S_1)_{ki}D_i,\quad\text{or}\nonumber\\
\sum_{i=1}^5 (S_2)_{ki}O_i &\ge\varepsilon-\sum_{i=1}^3 (S_1)_{ki}D_i,\label{eq:constr_crossing2}\\
k &=1 \dots 12.\nonumber
\end{align}

Therefore, the final frequency planning optimization problem at a fixed time $t$ can be posed as:
\begin{align}\label{eq:optimization_posed_final}
   & \quad\quad\quad\quad\quad\quad \underset{\vec{O}\in\mathbb{R}^5}{\operatorname{arg\, min}}\, || N_2\cdot\vec{O} - \vec{T} + N_1\cdot\vec{D}(t) ||^2\\
   & \begin{cases} 
        O_i \ge \min{}_i\,\nonumber\\
        -O_i \ge -\max{}_i\,,\quad i = 1,\dots, 5,\nonumber \\
        (M_2 \cdot \vec{O})_j \ge \min{}_j - (M_1 \cdot  \vec{D}(t))_j,\nonumber\\
        -(M_2 \cdot \vec{O})_j \ge - \max{}_j + (M_1 \cdot  \vec{D}(t))_j,\quad j = 1,\dots, 4,\nonumber\\
        (\sigma_C)_k(S_2\cdot\vec{O})_k \ge \varepsilon -(\sigma_C)_k (S_1\cdot\vec{D})_k,\quad k=1,\dots, 12.
    \end{cases}
\end{align}
This assumes the following as fixed: the Doppler shifts $\vec{D}(t)$ from the orbit, a feasible signs choice $(\sigma_O,\sigma_B)$ and a frequency range $f_{\min}-f_{\max}$ (to determine the $\min$ and $\max$ bounds of the constraints), target offset frequencies $\vec{T}$, and a given locking scheme to determine the matrices $M_1, M_2, S_1, S_2$.

By finding $\sigma_C$ so that this optimization reaches a feasible solution at every time of the mission duration, one obtains the offset frequencies time series $\vec{O}(t)$ and thus in turn the beatnote frequencies $\vec{B}(t)$. One can then search over the possible valid $\sigma_C$ that yield such complete solutions and keep the optimal offset frequencies for a given extra criterium, e.g., minimizing the beatnote frequency rate of change RMS.

The proposed algorithm has been implemented in C and in Python, and can be briefly summarized as follows:
\begin{enumerate}
    \item[\bf Step 1:] Fix a locking scheme from the possible configurations and obtain $M_1, M_2, S_1, S_2$.
    
    \item[\bf Step 2:] Compute the Doppler shifts $\vec{D}(t)$ from the orbit data.
    
    \item[\bf Step 3:] For a physically feasible frequency range $f_{\min}-f_{\max}$, and every possible sign choice $\sigma_O,\sigma_B$, determine hypercubes $\widehat{B}, \widehat{O}$, and compute the convex hull of the Minwkoski sum polytope $\widehat{B}\oplus -M_2\cdot\widehat{O}$.
    
    \item[\bf Step 4:] Compute $m(t,\sigma_B,\sigma_O)$, Eq. \eqref{eq:margin}, from the Minkowski sum face equations \eqref{eq:polytope_face}, for all times $t$ and every sign case of the previous step.
    
    \item[\bf Step 5:] From different ranges $f_{\min}-f_{\max}$, pick one such that $m_1>0$, Eq. \eqref{eq:margins}, fixing the corresponding $\sigma_O,\sigma_B$.
    
    \item[\bf Step 6:] Find the subset of sign choices $\sigma_C\in\Sigma_C$ compatible with the given $\sigma_O,\sigma_B$.
    
    \item[\bf Step 7:] For the frequency range $f_{\min}-f_{\max}$ fixed above, and each compatible $\sigma_C$ and time $t$, solve the optimization problem \eqref{eq:optimization_posed_final}.
    
    \item[\bf Step 8:] If more than one feasible solution $\vec{O}(t)$ is found for all times for different $\sigma_C$, choose the signs that yield time series satisfying an additional criterion, e.g., the beatnote frequencies have the smallest rate of change RMS, and keep those as final solutions.
\end{enumerate}

The method described above finds feasible time series for the offset frequencies which are nevertheless optimized with respect to target frequencies, $\vec{T}$, that are constant. This produces solutions that in practice may oscillate more than desired, since the target frequencies are not adapted at every time to match data from the orbit, in particular the target frequencies may not satisfy all the constraints of the system \eqref{eq:optimization_posed_final} at all times. Therefore, it is desirable to introduce a smoothing procedure.
\begin{figure}
\begin{centering}
\includegraphics[scale=0.45]{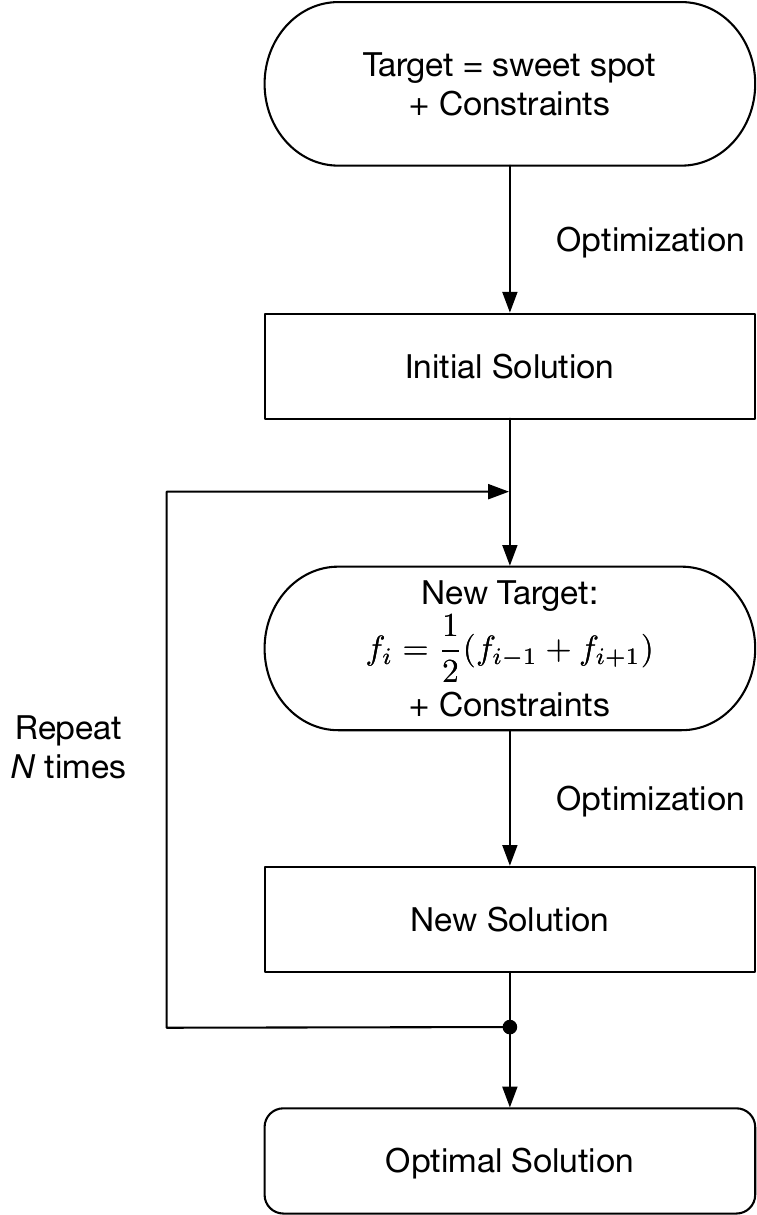}
\caption{ \label{fig:smoothing} Flowchart for the procedure of smoothing out the offset frequency time series solutions obtained from optimization.}
\end{centering}
\end{figure}

Once a frequency plan solution is obtained from the method above, one might be tempted to apply to each offset frequency time series a variety of smoothing filters, like averaging neighboring points, but the resulting smoothed-out series are no longer guaranteed to be feasible solutions to the optimization problem since at every instant they are certainly not going to satisfy the functional relation \eqref{eq:quad1} and the constraints, as the smoothing filters do not take those into account to produce their output. We thus propose to take these new filtered-out series as new target frequencies $\vec{T}(t)$, now time-varying, and run again the optimization process so that one obtains the closest feasible solutions to the smoothed-out ones. By repeating this procedure a number of iterations (see Fig. \ref{fig:smoothing}) the new offset frequency solutions are significantly smoothed-out while complying with the physical constraints and the dependency relations with the orbit data at every instant. We take this final output as our optimal frequency plan solution.

From these smooth solutions one can then apply interpolation methods that provide piece-wise interpolating polynomials between any two instants of time while preserving the values of the original solution and its first (discrete) derivative at the time nodes. The solutions found to the frequency planning are typically once per day, e.g., corresponding to the time resolution of the orbit data. An initial approach to obtain solutions at any instant of time would be to use piecewise linear interpolation, but the edges at midnight would not be optimal for TDI. Therefore we now have a postprocessing step to compute interpolating polynomials with restrictions not to cross constraint boundaries. A special spline algorithm has been implemented where we combine the PCHIP (piece-wise cubic Hermite interpolating polynomials) method with Bernstein polynomials to 5th order so that 1st and 2nd derivative of the spline are continuous and the spline does not over/under-swing the data-points. The coefficients of these interpolating polynomials can be periodically determined on-ground based on the latest orbit data, and sent via telecommand/telemetry to the onboard computer which then reprograms the interferometric detection system to update the phase-lock loop offsets.


\section{Results} \label{section:results}

\begin{figure}[t!]
\centering
\begin{tikzpicture}[font=\sffamily]
\definecolor{dkblue}{rgb} {0.00,0.33,0.68}
\definecolor{dkred}{rgb} {0.80,0.20,0.00}
\definecolor{lblue}{rgb} {0.00,0.60,0.95}
\definecolor{lred}{rgb} {0.95,0.30,0.00}
\tikzset{spacecraft/.style={circle, minimum width = 24mm, draw=black!60, fill=gray!10, very thick}}
\tikzset{laser/.style={rectangle, minimum width=10mm, minimum height=4mm, fill=red!20, draw=red!80, thick}}
\tikzset{masterlaser/.style={rectangle, minimum width=10mm, minimum height=4mm, fill=green!20, draw=green, thick}}
\tikzset{arrow/.style={line width=1mm}}
\tikzset{node distance = 0mm}
\coordinate (O) at (0,0); 
\coordinate (labelPosition) at (0,55mm); 
\draw[name path = sc1p, spacecraft] (0,0) circle [radius=\spacecraftRadius]
node (sc1) {};
\draw[name path = sc2p, spacecraft] (120:\spacecraftSeparation) circle [radius=\spacecraftRadius]
node (sc2) {};
\draw[name path = sc3p, spacecraft] (60:\spacecraftSeparation) circle [radius=\spacecraftRadius]
node (sc3) {};
\node (sc1l) at($(sc1.center)+(0,5mm)$) {\textbf{SC 1}};
\node (sc2l) at($(sc2.center)+(3mm,-1mm)$) {\textbf{SC 2}};
\node (sc3l) at($(sc3.center)+(-3mm,-1mm)$) {\textbf{SC 3}};
\coordinate (O1) at ($(sc1.center)+(0,-10mm)$); 
\coordinate (O2) at ($(sc2.center)+(150:10mm)$); 
\coordinate (O3) at ($(sc3.center)+(30:10mm)$); 
\path[name path = sc12] (O1) -- (O2);
\path[name path = sc13] (O1) -- (O3);
\path[name path = sc23] (O2) -- (O3);
\draw[name intersections={of=sc12 and sc1p, by=x}] node[laser, rotate=-60] (la12) at(x) {\laserOneTwo};
\draw[name intersections={of=sc12 and sc2p, by=x}] node[laser, rotate=-60] (la21) at(x) {\laserTwoOne};
\draw[name intersections={of=sc13 and sc1p, by=x}] node[laser, rotate=+60] (la13) at(x) {\laserOneThree};
\draw[name intersections={of=sc13 and sc3p, by=x}] node[laser, rotate=+60] (la31) at(x) {\laserThreeOne};
\draw[name intersections={of=sc23 and sc2p, by=x}] node[laser, rotate=  0] (la23) at(x) {\laserTwoThree};
\draw[name intersections={of=sc23 and sc3p, by=x}] node[masterlaser, rotate=  0] (la32) at(x) {\laserThreeTwo};
\draw[latex-, arrow, lblue] (la12)--(la21) node[midway, above=2mm, sloped, black] {$L_{21} + D_3$};
\draw[latex-, arrow, lblue] (la23)--(la32) node[midway, above=2mm, sloped, black] {$L_{32} + D_1$};
\draw[-latex, arrow, lblue] (la12.east)..controls +(0.75,-1) and +(-0.75,-1)..(la13.west);
\draw[latex-, arrow, lblue] (la21.west)..controls +(-0.5,1.20) and +(-1.20,0.0)..(la23.west);
\draw[latex-, arrow, lred] (la31.east)..controls +(0.5,1.20) and +(1.20,0.0)..(la32.east);
\end{tikzpicture}
\caption{Constellation configuration for the non-swap locking scheme N3-L32.}
\label{fig:constellation}
\end{figure}
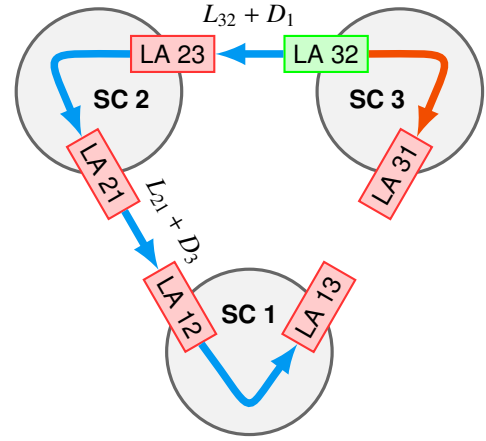



We shall now illustrate the general mathematical method described in the previous section to generate a particular frequency plan for the locking configuration N3-L32, Figure \ref{fig:constellation}, given the orbit data \texttt{'op25-340-10y3t.xyz'} provided by ESA. This orbit time series spans the whole mission duration of 10 years.

\textbf{Step 1:} The first step is fixing the locking scheme which, in this particular case, has the laser frequency $L_{32}$ set as primary and where the other transponder lasers are locked according to:
\begin{align}
& L_{32} \text{ (primary)},\nonumber \\
& L_{23} = L_{32} + O_1 + D_1,\nonumber \\
& L_{13} = L_{12} + O_2,\nonumber \\
& L_{31} = L_{32} + O_3,\nonumber \\
& L_{21} = L_{23} + O_4,\nonumber \\
& L_{12} = L_{21} + O_5 + D_3.
\label{eq:lock_initial}
\end{align}
This is a valid locking scheme where the linear system of dependencies has a unique solution, i.e., all transponder laser frequencies can be solved for, and thus referenced to the primary laser frequency, leading to:
\begin{align}
& L_{32} \text{ (primary)} \nonumber \\
& L_{23} = L_{32} + D_1 + O_1, \nonumber \\
& L_{13} = L_{32} +D_1 + D_3 + O_1 + O_2 + O_4 + O_5, \nonumber \\
& L_{31} = L_{32} + O_3, \nonumber \\
& L_{21} = L_{32} +D_1 + O_1 + O_4, \nonumber \\
& L_{12} = L_{32} +D_1 + D_3 + O_1 + O_4 + O_5.
\label{eq:lock_final}
\end{align}
The beatnote frequencies for this non-swap case are given by \eqref{eq:beatn} and thus, in terms of the offset frequencies and Doppler shifts using \eqref{eq:lock_final}, they become:
\begin{alignat}{2}
B_{11}  &=  O_2                                                \nonumber \\
B_{12}  &=  - O_5                                           \nonumber \\
B_{13}  &=  - D_1 + D_2 - D_3 - O_1 - O_2 + O_3 - O_4 - O_5        \nonumber \\
B_{21}  &=  2 D_3 + O_5                                    \nonumber \\
B_{22}  &=  O_4                                                \nonumber \\
B_{23}  &=  - O_1                                           \nonumber \\
B_{31}  &=  D_1 + D_2 + D_3 + O_1 + O_2 - O_3 + O_4 + O_5        \nonumber \\
B_{32}  &=  2 D_1 + O_1                                    \nonumber \\
B_{33}  &=  - O_3                                                
\label{eq:beatn2}
\end{alignat}
This has the following matrix form according to the notation of Equation \eqref{eq:quad1}:

\begin{align}
\begin{array}{c}
 \left(
  \begin{array}{c}
B_{23}\\
B_{11}\\
B_{33}\\
B_{22}\\
B_{12}
  \end{array}
 \right)\\
 \left(
  \begin{array}{c}
B_{13}\\
B_{21}\\
B_{31}\\
B_{32}
  \end{array}
 \right)
\end{array}
\!\!=
\left(
\begin{array}{cc}
 \left(
  \begin{array}{ccc}
       0 & 0 & 0     \\   
       0 & 0 & 0     \\
       0 & 0 & 0     \\
       0 & 0 & 0     \\
       0 & 0 & 0     
  \end{array}
 \right)
&
 \left(
  \begin{array}{ccccc}
         -1  &   0 &    0   &  0 &    0    \\   
          0  &   1 &    0   &  0 &    0    \\
          0  &   0 &   -1   &  0 &    0    \\
          0  &   0 &    0   &  1 &    0    \\
          0  &   0 &    0   &  0 &   -1    
  \end{array}
 \right)\\
\underbrace{
 \left(
  \begin{array}{ccc}
      -1 & 1 &-1     \\
       0 & 0 & 2     \\
       1 & 1 & 1     \\
       2 & 0 & 0     
  \end{array}
 \right)}_{N_1}
&
\underbrace{
 \left(
  \begin{array}{ccccc}
       -1  &  -1 &    1   & -1 &   -1    \\
        0  &   0 &    0   &  0 &    1    \\
        1  &   1 &   -1   &  1 &    1    \\
        1  &   0 &    0   &  0 &    0    
  \end{array}
 \right)}_{N_2}
 \end{array}
 \right)
\!\cdot\!\!
\begin{array}{c}
 \left(
  \begin{array}{c}
D_1 \\
D_2 \\
D_3 
  \end{array}
 \right)\\
 \left(
  \begin{array}{c}
O_1 \\
O_2 \\
O_3 \\
O_4 \\
O_5 
  \end{array}
 \right)
\end{array}
\label{eq:m1}
\end{align}
Here the components have been reordered so that the first 5 entries are the locking beatnotes sorted to yield a diagonal block, and the last 4 components are the non-locking ones in lexicographical order. This determines the matrices $N_1$ and $N_2$, and $M_1, M_2$ by removing the first 5 rows.
 \begin{figure*}[t!]
    \centering
    \includegraphics[width=\textwidth]{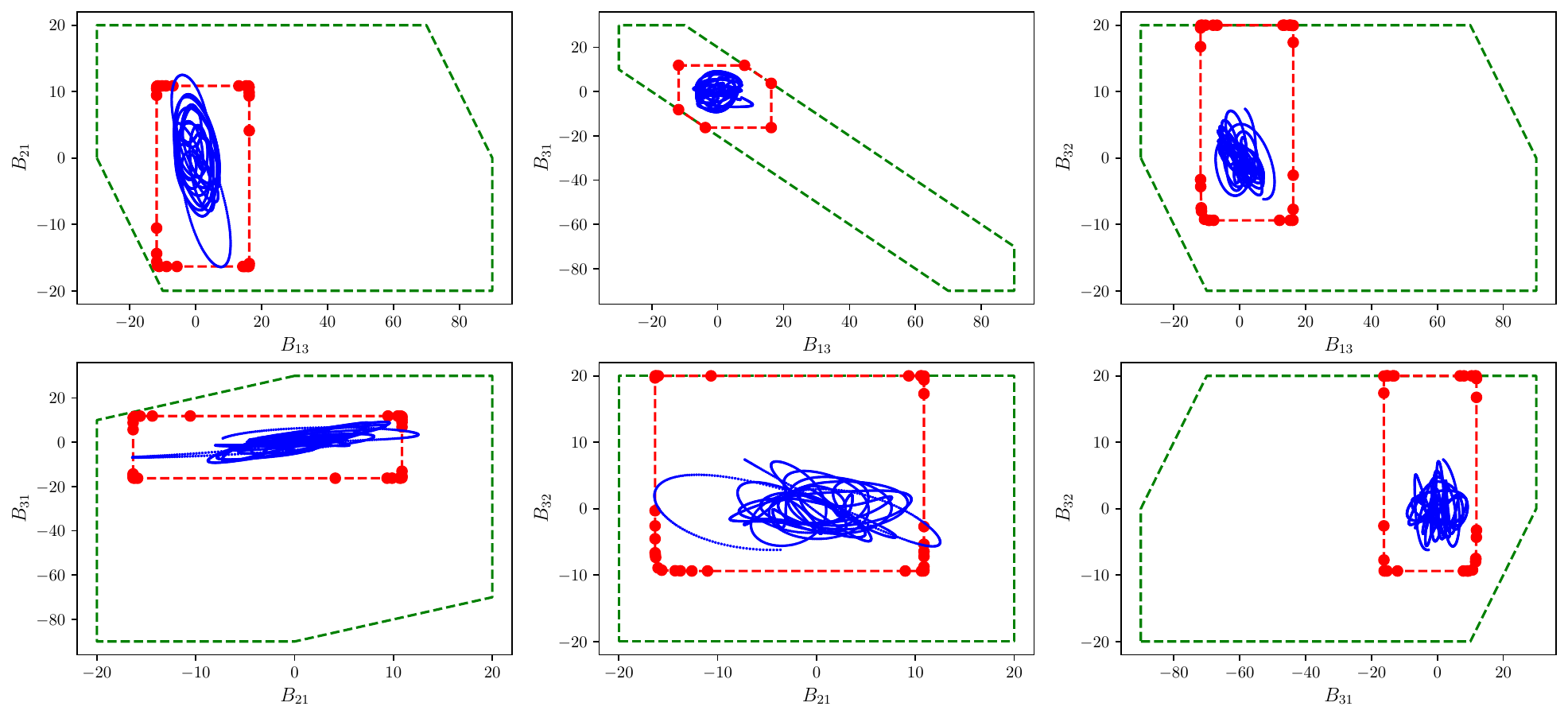}
    \caption{Constraining volumes in the beatnote frequency space for configuration N3-L32 at day 102 of the orbits from \texttt{`op25\_340\_10y3t.xyz'}. The green boundary represents the convex hull for a feasible frequency band and beatnote sign choices since it completely encloses the blue orbit data. The red boundary represents the outer boundary of the full set of constraints at the initial time, and it must enclose the corresponding orbit point only at that time.} 
    \label{fig:beatnote_polytope}
\end{figure*}

In turn, the differences and sums of beatnote frequencies appearing in the non-crossing constraints of Equation \eqref{eq:DB_SB} are
\begin{align}
& \Delta B_1 =  B_{12}-B_{11},\nonumber\\
&  \Delta B_2 = B_{13}-B_{11},\nonumber\\
&  \Delta B_3 = B_{21}-B_{22},\nonumber\\
&  \Delta B_4 = B_{23}-B_{22},\nonumber\\
& \Delta B_5 =  B_{31}-B_{33},\nonumber\\
& \Delta B_6 =  B_{32}-B_{33},\nonumber\\
& \Delta B_7 =  B_{12}+B_{11},\nonumber\\
& \Delta B_8 =  B_{13}+B_{11},\nonumber\\
& \Delta B_9 =  B_{21}+B_{22},\nonumber\\
& \Delta B_{10} =  B_{23}+B_{22},\nonumber\\
& \Delta B_{11} =  B_{31}+B_{33},\nonumber\\
& \Delta B_{12} =  B_{32}+B_{33},
\end{align}
which yield the following matrices $S_1, S_2$ from the relations \eqref{eq:m1}:
\begin{align}
\begin{pmatrix}
\Delta B_1\\
\Delta B_2\\
\Delta B_3\\
\Delta B_4\\
\Delta B_5\\
\Delta B_6\\
\Delta B_7\\
\Delta B_8\\
\Delta B_9\\
\Delta B_{10}\\
\Delta B_{11}\\
\Delta B_{12}
\end{pmatrix}
=         
\underbrace{
 \left(
\begin{array}{ccc}
       0 & 0 & 0    \\   
      -1 & 1 & -1   \\
       0 & 0 & 2     \\
       0 & 0 & 0    \\
      1 & 1 & 1   \\
      2 & 0 & 0    \\
      0 & 0 & 0    \\
      -1 & 1 & -1    \\
       0 & 0 & 2    \\
       0 & 0 & 0 \\
       1 & 1 & 1\\
       2 & 0 & 0
\end{array}
\right)}_{S_1}
 \cdot
\begin{pmatrix}
D_1 \\
D_2 \\
D_3
\end{pmatrix}
+
\underbrace{
 \left(
\begin{array}{ccccc}
       0  &   -1 &    0   &  0 &    -1   \\   
      -1  &   -2 &    1   &  -1 &    -1    \\
       0  &   0 &   0   &  -1 &    1    \\
      -1  &   0 &    0   &  -1 &    0    \\
       1  &   1 &    0   &  1 &   1    \\
      1  &  0 &  1   & 0 &   0    \\
      0  &   1 &    0   &  0 &    -1    \\
       -1  &  0 &   1   &  -1 &    -1    \\
       0  &   0 &    0   &  1 &    1    \\
       -1 & 0 & 0 & 1 & 0 \\
       1 & 1 & -2 & 1 & 1 \\
       1 & 0 & -1 & 0 & 0
\end{array}
\right)}_{S_2}
 \cdot
\begin{pmatrix}
O_1 \\
O_2 \\
O_3 \\
O_4 \\
O_5 
\end{pmatrix}
\label{eq:mat_DBSB}
\end{align}

\begin{figure*}[t!]
    \centering
    \includegraphics[width=\textwidth]{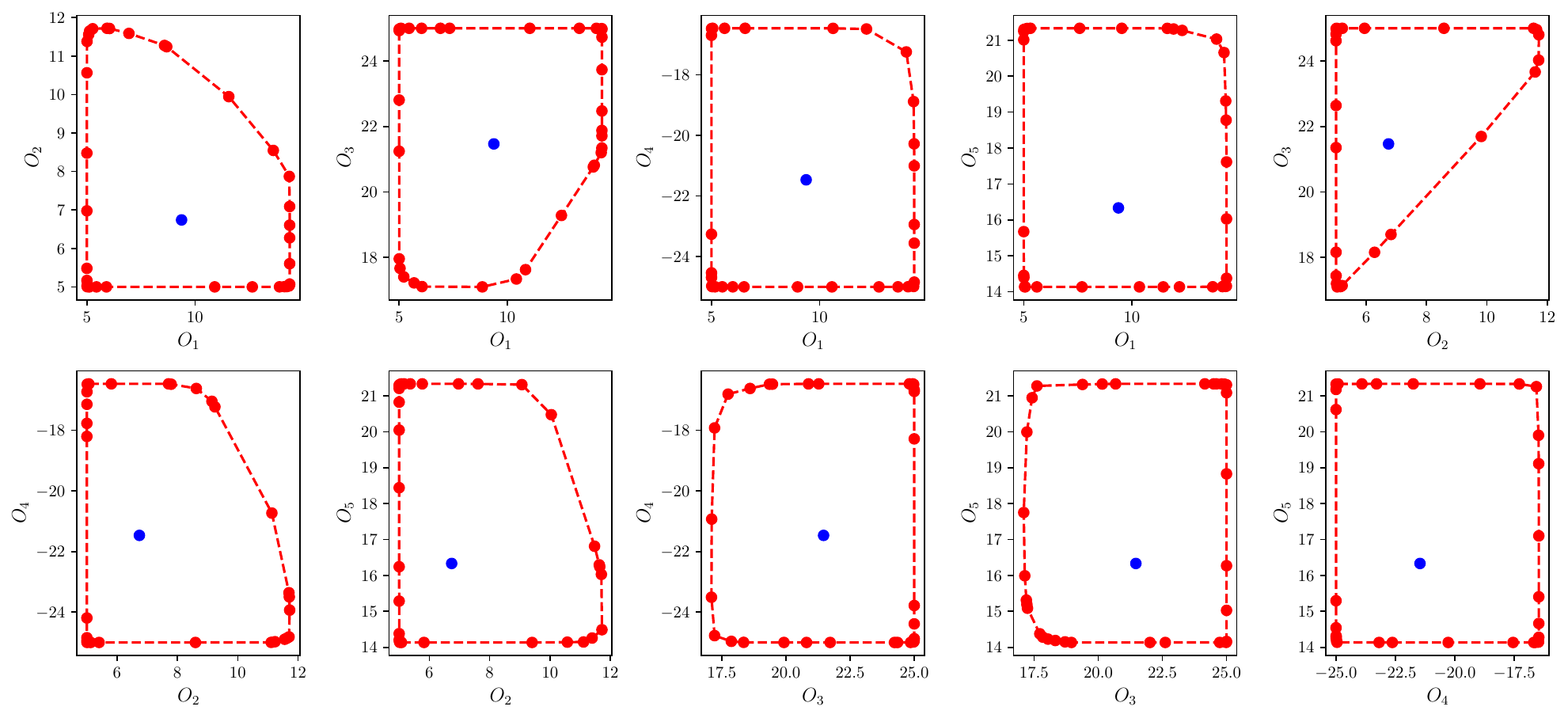}
    \caption{Constraining domain outer boundaries (in red) for the offset frequencies for configuration N3-L32 at day 102 of the orbits from \texttt{`op25\_340\_10y3t.xyz'}.}
    \label{fig:offset_polytope}
\end{figure*}

\textbf{Step 2:} The second step is to fix the values of the Doppler shifts $D_1(t), D_2(t), D_3(t)$ at a given time $t$, obtained from this example's orbit data, as in Figure~\ref{figure:constellation}.

\textbf{Steps 3/4:} Next, let us mention one of the possible cases that the algorithm must explore, and which happens to be the actual final optimal choice the method achieves for this example: fix the feasible frequency range to
\begin{equation}
    f_{\min} = 5\,\text{MHz},\quad f_{\max} = 25\,\text{MHz},
\end{equation}
and the offset frequency signs and the non-locking beatnote frequency signs to
\begin{equation}
    \sigma_O = (1, 1, 1, -1, 1),\quad  \sigma_B=(1, 1, -1, 1).
\end{equation}
This means we are considering the following 5-dimensional hypercube domain $\widehat{O}$ arising from the frequency ranges of every component of the offset vector:
\begin{align}\label{eq:cubeO}
    \widehat{O} = \{ \vec{O}\in\mathbb{R}^5\;\vert\; 
                    5 &\le O_{1} \le 25, \nonumber\\
                    5 &\le O_{2} \le 25,  \nonumber\\
                    5 &\le O_{3} \le 25, \nonumber\\
                    -25 &\le O_{4} \le -5, \nonumber\\
                    5 &\le O_{5} \le 25 \}.
\end{align}
Similarly, the 4-dimensional hypercube domain $\widehat{B}$ from the frequency ranges of every component of the non-locking beatnote vector is:
\begin{align}\label{eq:cubeB}
    \widehat{B} = \{ \vec{B}_{\rm non-lock}\in\mathbb{R}^4\;\vert\; 
                    5 &\le B_{13} \le 25, \nonumber\\
                    5 &\le B_{21} \le 25,  \nonumber\\
                    -25 &\le B_{31} \le -5, \nonumber\\
                    5 &\le B_{32} \le 25 \}.
\end{align}
The vertices or corners of $\widehat{O}$ are all the possible combinations of the boundary values of the components $O_i,i=1,\dots,5$, taking into account the sign choices, i.e., the set of $2^5$ vectors
\begin{equation}
\widehat{O}_C := \{ \vec{O}\in\mathbb{R}^5\,\vert\, O_i\in\{(\sigma_O)_i f_{\rm min}, (\sigma_O)_i f_{\max}\},i =1,\dots, 5 \}
\end{equation}
for example $(5,25,5,-25,25)$. This is a finite set whose linear image through the map $M_2:\mathbb{R}^5\rightarrow\mathbb{R}^4$ is easily computed by matrix multiplication. A similar set specifies the $2^4$ vertices of $\widehat{B}$ for the non-locking beatnote components:
\begin{equation}
\widehat{B}_C := \{ \vec{B}\in\mathbb{R}^4\,\vert\, B_i\in\{(\sigma_B)_i f_{\rm min}, (\sigma_B)_i f_{\max}\},i =1,\dots, 4 \}.
\end{equation}
Since both $-M_2\cdot\widehat{O}_C$ and $\widehat{B}_C$ are comprised of finitely many vectors in $\mathbb{R}^4$, the Minkowski sum $\widehat{B}_C\oplus (-M_2\cdot\widehat{O}_C)$ is straightforwardly computed by performing all possible linear combinations between them, e.g., 
\begin{equation}
\begin{pmatrix}5\\ 25\\ -5\\ 5\end{pmatrix}+ M_2\cdot\begin{pmatrix}5\\ 25 \\ 5 \\ -25 \\ 25\end{pmatrix}=\begin{pmatrix}
    -20\\ 50\\ 20\\ 10 \end{pmatrix}  \in \widehat{B}_C\oplus (-M_2\cdot\widehat{O}_C).
\end{equation}
Then, through a software package like \texttt{qhull}, we determine a complete representative of the convex hull of the actual convex polytope $\widehat{B}\oplus (-M_2\cdot\widehat{O})$ by computing the convex hull of $\widehat{B}_C\oplus (-M_2\cdot\widehat{O}_C)$. This provides us with hyperplane equations for all faces, and thus, all the data needed to evaluate the margin function $m(t,\sigma_B,\sigma_O)$, Equation \eqref{eq:margin}, for the specific Doppler shifts vector $\vec{x}=M_1\cdot\vec{D}(t)$.

The green dashed boundary of Figure~\ref{fig:offset_polytope} represents the projection to every coordinate plane (in the beatnote frequency space $\mathbb{R}^4$, where the sums of $\widehat{B}\oplus(-M_2\cdot\widehat{O})$ happen) of the Minkowski sum boundary just computed. The frequency band and sign choices are feasible because the orbit data, represented by the vectors $M_1\cdot\vec{D}(t)$, stays inside this convex domain.

The sign choices $\sigma_O = (1, 1, 1, -1, 1)$ and $\sigma_B=(1, 1, -1, 1)$ were determined by repeating the above procedure for all sign combinations, computing the function $m_1$ from Equation \eqref{eq:margins}, in other words, they are the outcome of: $$\underset{\sigma_O\in\Sigma_O,\;\sigma_B\in\Sigma_B}{\operatorname{arg}\operatorname{max}}\,\underset{t}{\operatorname{min}}\; m(t,\sigma_B,\sigma_O),$$ which checks whether $\vec{x}(t)=M_1\cdot\vec{D}(t)$ lies inside the corresponding convex hull of the Minkowski sum as computed above. Exploring different frequency bands yields Figure~\ref{fig:contours}, which confirms that our choice  of 5\,MHz--25\,MHz for this example lies in Case 3, i.e. $m_1>0$ as explained in Subsection \ref{subsec:geometry_signs}, and so this band guarantees that these $\sigma_O,\sigma_B$ are constant and do not need to be updated throughout the whole mission duration for this orbit time series, avoiding interruptions in the science streams. This solves step 5 of the method.

\textbf{Step 5:} We fix $\fmin = 5\,$MHz and $\fmax = 25\,$MHz, as per the previous steps. 

\textbf{Step 6:} In this step we include additional constraint sign possibilities, leading to the final choice of 
\begin{equation}\label{eq:sigmaC_ex}
    \sigma_C=(-1, 1, 1, 1, 1, 1, -1, 1, -1, -1, -1, -1)
\end{equation}
for this example. This is done by reducing the 4096 possible cases to those that are consistent with the already fixed $\sigma_O = (1, 1, 1, -1, 1), \sigma_B=(1, 1, -1, 1)$. For example, looking at $\Delta B_1 = B_{12}-B_{11}$, with $B_{12}= -O_5$ and $B_{11}= +O_2$ by Eq. \eqref{eq:m1}, we see that $\Delta B_1 \leq -10$, since $O_2 \geq +5, O_5\geq +5$ per the values of $\sigma_O$, and therefore there is no consistent choice possible for $(\sigma_C)_1$ but $-1$. The components where this sign determination can not be concluded in this manner must be explored in the optimization phase in order to choose the best results (step 8).

\textbf{Step 7:} This step is the core of the method, as it performs the optimization process. This consists of finding offset frequency values $\vec{O}$ that minimize $|| N_2\cdot\vec{O} - \vec{T} + N_1\cdot\vec{D}(t) ||^2$, where the initial target frequencies are constant and fixed to the average of $f_{\min}$ and $f_{\max}$,
\begin{equation}
    \vec{T} = (-15,15,-15,-15,-15, 15, 15, -15, 15),
\end{equation}
(notice the consistency of the last 4 components with the signs of $\sigma_B$, and the first 5 components with $\sigma_O$ acted upon by the upper block of $N_2$). The explicit function to minimize is the quadratic polynomial
\begin{align}
& || N_2\cdot\vec{O} - \vec{T} + N_1\cdot\vec{D}(t) ||^2 = \nonumber\\
  & 8 D_1 O_1+4 D_1 O_2-4 D_1 O_3+4 D_1 O_4+4 D_1 O_5 +\nonumber \\
  & 4 D_3 O_1+4 D_3 O_2-4 D_3 O_3+4 D_3 O_4+8 D_3 O_5+\nonumber \\
  &6 D_1^2+4 D_3 D_1+2 D_2^2+6 D_3^2+4 O_1^2+3 O_2^2+\nonumber \\
  &3 O_3^2+3 O_4^2+4 O_5^2+4 O_1 O_2+30 O_2-4 O_1 O_3+\nonumber \\
  &-4 O_2 O_3-90 O_3+4 O_1 O_4+4 O_2 O_4-4 O_3 O_4+\nonumber \\
  &90 O_4+4 O_1 O_5+4 O_2 O_5-4 O_3 O_5+4 O_4 O_5+2025,
\end{align}
where the Doppler shifts are considered here as constant coefficients (fixed at a given time) and the offset frequencies are the independent variables.

The first 10 constraints of \eqref{eq:optimization_posed_final} come from the offset frequency band and are those of \eqref{eq:cubeO}, whereas the next 8 constraints arise from the non-locking beatnote frequency band, i.e., those substituting $\vec{B}_{\rm non-lock}$ by $M_1\cdot\vec{D}+M_2\cdot\vec{O}$ in \eqref{eq:cubeB}. Finally, the remaining 12 constraints arise from \eqref{eq:mat_DBSB} via $(\sigma_C)_k\Delta B_k \geq \varepsilon$ with $\sigma_C$ given by \eqref{eq:sigmaC_ex}, and $\varepsilon=2\,\text{MHz}$ the non-crossing margin for this example.

\begin{figure*}[t!]
\begin{subfigure}{0.49\linewidth}
\centering
\includegraphics[width=\linewidth]{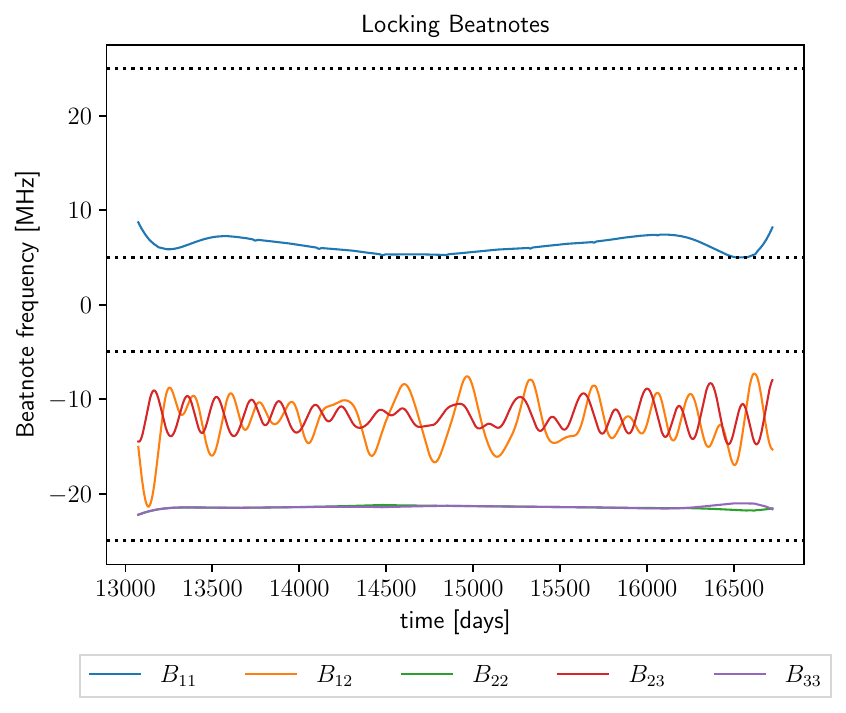}
\caption{Optimized beatnotes for configuration N3-L32.}
\label{fig:beatnotes_lock_time series}
\end{subfigure}\hfill
\begin{subfigure}{0.49\linewidth}
\centering
\includegraphics[width=\linewidth]{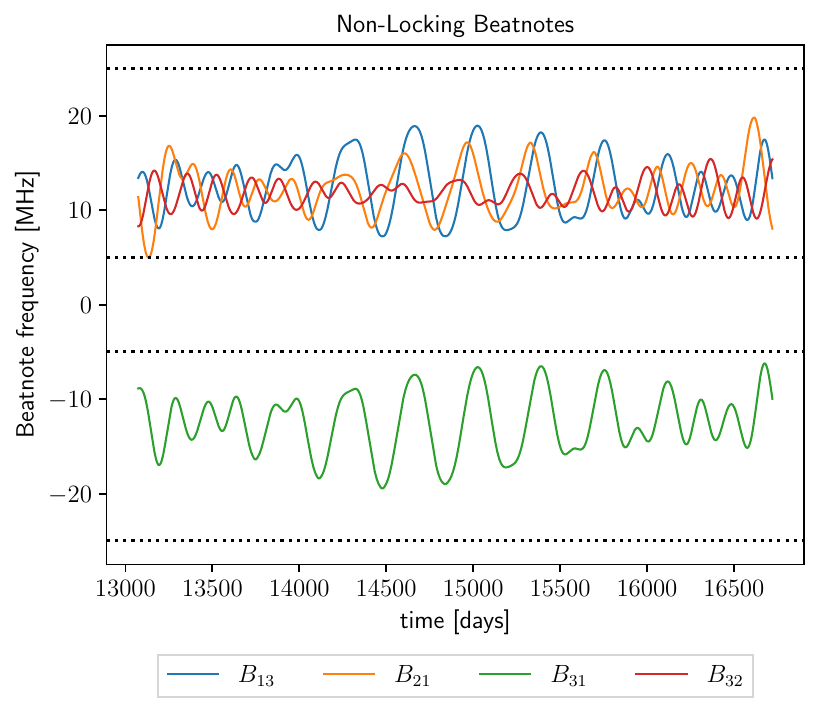}
\caption{Optimized non-locking beatnotes for configuration N3-L32.}
\label{fig:beatnotes_nonlock_time series}
\end{subfigure}\hfill
\caption{}
\label{}
\end{figure*}

Taking into account all the 30 constraints reduces the feasible solution domain for the offset frequencies, so that it is no longer the hypercube $\widehat{O}$ but a non-necessarily convex polytope (i.e., a finite region bounded by flat boundaries that may contain holes). The outer boundary of this fully constrained domain can be nevertheless seen in Figure \ref{fig:offset_polytope}, where the red dashed lines represent the outer boundaries of this region projected onto the coordinate planes of the offset space $\vec{O}\in\mathbb{R}^5$. The blue dot represents valid offset frequencies that are a feasible solution to the optimization problem with all the parameters as stated above. By projecting this region to the non-locking beatnote frequency space $\vec{B}\in\mathbb{R}^4$ through matrix $M_2$, we can see in Figure \ref{fig:beatnote_polytope} the outer boundary of the constraining polytope surrounding the orbit data, represented in this space by $M_1\cdot \vec{D}(t)$, which for different times yields the blue curve. If this latter vector were to lie outside the red-bounded region the sign choices and frequency band would not provide a feasible choice to run the optimization.

The minimization process for every time $t$ of the orbit time series outputs the final offset frequencies which, after the smoothing explained at the end of the previous section, results in the beatnotes time series of Figure \ref{fig:beatnotes_lock_time series}. The method employed to solve the optimization of this example was \texttt{SciPy}'s Sequential Least Squares Programming (SLSQP). We can immediately appreciate that the optimization in this example makes the local beatnotes $B_{11}, B_{22}, B_{33}$ almost constant. 

\begin{figure*}
\begin{subfigure}{0.49\linewidth}
\centering
\includegraphics[width=\linewidth]{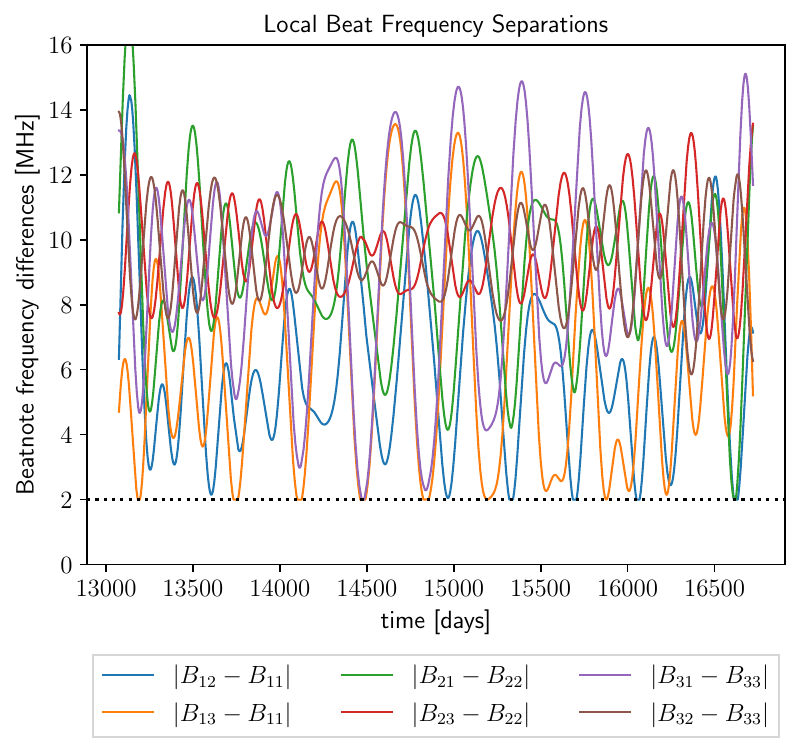}
\caption{Difference between beatnotes for configuration N3-L32.}
\label{fig:beatnotes_differences}
\end{subfigure}\hfill
\begin{subfigure}{0.49\linewidth}
\centering
\includegraphics[width=\linewidth]{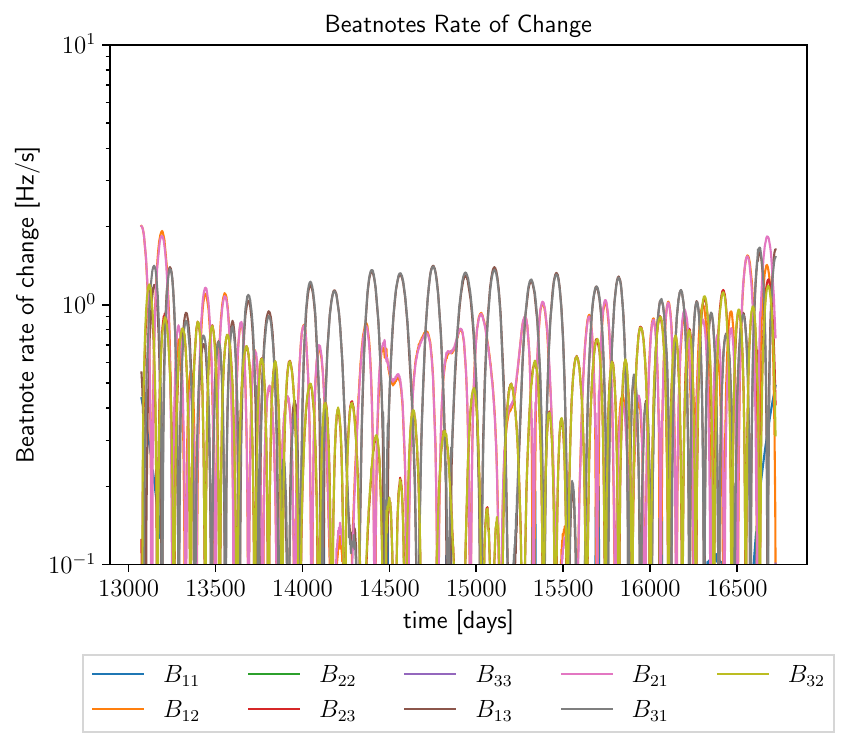}
\caption{Rate of change of beatnotes for configuration N3-L32.}
\label{fig:beatnotes_rates}
\end{subfigure}\hfill
\caption{}
\end{figure*}

The frequency band margins of 5--25\,MHz are also shown in dotted lines, confirming that the solutions indeed lie inside the band, taking into account the signs specified by $\sigma_B$ for the non-locking beatnotes, and the signs of the locking beatnotes given by the upper block of $N_2$ acting on $\sigma_O$. The beatnote differences $|\,|B_{ij}|-|B_{ii}|\,|$ are shown in Figure \ref{fig:beatnotes_differences}, where the non-crossing threshold of 2\,MHz was chosen, confirming the solutions also satisfy these constraints. 

\textbf{Step 8:} Finally, the sign $\sigma_C$ is optimized to minimize the RMS of the rates of change of beatnote frequencies (Figure~\ref{fig:beatnotes_rates}).

Once this initial solution is found, $\vec{O}_{in}(t)$, and all the sign choices have been made, we repeat the optimization process using as new target frequencies the smoothed-out time series of the initial solution, see Figure \ref{fig:smoothing}, i.e., $\vec{T}(t)=\vec{B}_{in}(t)$, with the beatnotes $\vec{B}_{in}(t)$ those corresponding to the solution $\vec{O}_{in}(t)$. The final output thus consists of the smoother time series $\vec{O}_{out}(t)$.

These solutions provide values for the offset frequencies for every day of the orbit time. In this example, for simplicity, we compute the piece-wise cubic Hermite polynomials that interpolate between each pair of days to produce offset values at a much finer time resolution, while maintaining the exact values of the original solution (and its discrete first derivatives) at the daily time breakpoints, and avoiding overshooting of the solution beyond the bandwidth constraint boundaries. For instance, during the first two days (i.e. for $t$ in between $t_0=13074.1406279,$ and $t_1=13075.1406279$), we obtain the following polynomials:
\begin{align}
O_1(t) =  
-0.0006532907752644507\;(t - t_0)^3\; & +\nonumber\\ 
-0.001690277532664028\;(t - t_0)^2\; & +\nonumber \\
0.01309428489022757\;(t - t_0)\; & + \nonumber\\
14.483126147178096, &
\end{align}
\begin{align}
O_2(t) =  
7.538946707463801\cdot 10^{-7}\;(t - t_0)^3\; & + \nonumber\\ 
 0.00016748847683473178\;(t - t_0)^2\; & +\nonumber \\
-0.03788217470242561\;(t - t_0)\; & + \nonumber\\
8.71039204492068, &
\end{align}
\begin{align}
O_3(t) =  
3.822182343196645\cdot 10^{-9}\;(t - t_0)^3\; & + \nonumber\\ 
 5.212005644189208\cdot 10^{-6}\;(t - t_0)^2\; & +\nonumber \\
-0.007128056477036182\;(t - t_0)\; & + \nonumber\\
22.227273698100003, &
\end{align}
\begin{align}
O_4(t) =  
-8.747618260637052\cdot 10^{-7}\;(t - t_0)^3\; & + \nonumber\\ 
-7.846611974351471\cdot 10^{-5}\;(t - t_0)^2\; & +\nonumber \\
0.007354898265365506\;(t - t_0)\; & + \nonumber\\
-22.227541666452975, &
\end{align}
\begin{align}
O_5(t) =  
 -3.0812613638353525\cdot 10^{-7}\;(t - t_0)^3\; & + \nonumber\\ 
-0.00023074876614948003\;(t - t_0)^2\; & +\nonumber \\
0.17372650511784826\;(t - t_0)\; & + \nonumber\\
15.043307304145314. &
\end{align}

Coefficients such as these for 5th order PCHIP-Bernstein splines are the actual frequency plan data that must be periodically delivered to the onboard computers for reconfiguration of the corresponding phasemeters in the constellation such that they tune the phase-lock loop offsets accordingly.


\section{Conclusion} \label{section:conclusion}

In this paper, we have described a computational-geometric method, based on the Minkowski sum of convex polytopes, to study the valid heterodyne frequency ranges that avoid the need to switch the signs of the laser transponder offsets throughout the lifetime of the LISA mission. Choosing a frequency range $(\fmin,\,\fmax)$ belonging to ``Case 3'' described in Section~\ref{subsec:geometry_signs} (Figure~\ref{fig:contours}) guarantees that the laser locks can be maintained without interruption for at least 10 years, maximizing the uptime of the LISA detector and thus the extent of the science data stream.

This has also allowed us to pose the physical constraints, given by the feasible frequency bands of each offset and beatnote and the local non-crossing requirements, in terms of linear inequality constraints that cut out those geometric domains. Informed by this geometric interpretation, we have then proposed a quadratic programming minimization algorithm to find offset frequencies that make the beatnotes as close as possible to certain constant target frequencies, that are then adapted at every time to find a solution as smooth as possible.

This furnishes us with cubic piece-wise polynomials between any two successive instants of the orbit data that interpolate the offset frequencies. The coefficients of these polynomials, computed periodically on-ground based on the latest orbit data, is the final payload to be transmitted to the onboard computers in the constellation for in-flight implementation of the frequency plan.

The methods and algorithms described in this paper are not exclusive to LISA, but can be applied to other missions employing similar laser transponder schemes.

\section{References}
%


\section*{Acknowledgements}

The authors acknowledge financial support by the German Aerospace Center (DLR) with funds from the Federal Ministry of Economics and Technology (BMWi) according to a decision of the German Federal Parliament (Grants No.\ 50OQ0601, No.\ 50OQ1301, No.\ 50OQ1801), the European Space Agency (ESA) (Grants No.\ 22331/09/ NL/HB, No.\ 16238/10/NL/HB), the Deutsche Forschungsgemeinschaft (DFG) Sonderforschungsbereich 1128 Relativistic Geodesy and Cluster of Excellence ``QuantumFrontiers: Light and Matter at the Quantum Frontier: Foundations and Applications in Metrology'' (EXC-2123, Project No.\ 390837967).

\end{document}